\documentclass[useAMS,usenatbib,letterpaper]{mn2e}
\hypersetup{draft}
\usepackage{amsmath}
\usepackage{txfonts}
\usepackage{graphics}
\usepackage{times}
\usepackage{todonotes}
\usepackage{bm}

\renewcommand{\vec}{\bm}
\newcommand{\vechat}[1]{\bm{\hat{#1}}}

\begin{document}


\title[Matched filtering with $21\,$cm]{Matched filtering with
interferometric $21\,$cm experiments}
\author[White and Padmanabhan]{
  Martin White$^{1,2,3}$, Nikhil Padmanabhan$^{4}$ \\
  $^{1}$ Department of Astronomy, University of California,
  Berkeley, CA 94720, USA \\
  $^{2}$ Department of Physics, University of California,
  Berkeley, CA 94720, USA \\
  $^{3}$ Lawrence Berkeley National Laboratory, 1 Cyclotron Road,
  Berkeley, CA 94720, USA \\
  $^{4}$ Department of Physics, Yale University, New Haven, CT 06511, USA.
}
\date{\today}
\pagerange{\pageref{firstpage}--\pageref{lastpage}}

\maketitle

\label{firstpage}

\begin{abstract}
A new generation of interferometric instruments is emerging which aim to
use intensity mapping of redshifted $21\,$cm radiation to measure the
large-scale structure of the Universe at $z\simeq 1-2$ over wide areas of sky.
While these instruments typically have limited angular resolution, they
cover huge volumes and thus can be used to provide large samples of 
rare objects.  In this paper we study how well such instruments could find
spatially extended large-scale structures, such as cosmic voids, using a
matched filter formalism.  Such a formalism allows us to work in Fourier
space, the natural space for interferometers, and to study the impact of
finite $u-v$ coverage, noise and foregrounds on our ability to recover
voids.  We find that, in the absence of foregrounds such, instruments would
provide enormous catalogs of voids, with high completeness, but that
control of foregrounds is key to realizing this goal.
\end{abstract}



\section{Introduction}
\label{sec:intro}

Intensity mapping with radio interferometers has emerged as a potentially powerful means
of efficiently mapping large volumes of the Universe, albeit at low
spatial resolution.  Several groups are fielding $21\,$cm intensity
mapping experiments using a variety of technical designs
\citep{BAOradio12,Tianlai12,BAObab13,CHIME14,ORT14,HIRAX}.
Even though such instruments do not have the angular resolution to
see individual galaxies, or even large clusters, they are capable of
mapping the larger elements of the cosmic web (e.g.~protoclusters and
cosmic voids).
Protoclusters are the progenitors of the most massive systems in the
Universe today \citep{Ove16}.
Cosmic voids make up most of the Universe, by volume \citep{Roo88,Wey11}.
As we shall show, in each case the system size is sufficiently large that
they can be reliably found with upcoming intensity mapping experiments if
foregrounds can be controlled sufficiently well.

Cosmic voids, regions almost devoid of galaxies, are intrinsically interesting
as the major constituent of the cosmic web by volume, and as an extreme
environment for galaxy evolution \citep{Roo88,Wey11}.
They may be an excellent laboratory for studying material that clusters
weakly like dark energy \citep{LeePar09,LavWan12}
or neutrinos \citep{BanDal16})
or for testing modified gravity
\citep{Cla13,Ham14,Ham15,Cai15,Cai16,Fal17,Ham17}.
In this paper we show that $21\,$cm instruments aimed at measuring the
large-scale power spectrum, either proposed or under construction, should
enable a search of enormous cosmic volumes at high redshift for these rare
objects, which are large enough to be detected at high significance
\citep[see also][]{Bat04}.

Voids and protoclusters are inherently ``configuration space objects'', in
the sense of being highly coherent under- or overdensities in the matter
field in configuration space.  However most future $21\,$cm experiments are
interferometers which naturally work in Fourier space.  We will use a
matched filter formalism to allow us to work in the interferometer's natural
space, where the noise and sampling are easy to understand.  This formalism
also provides a natural way to combine data sets which live in different
domains, e.g.~optical imaging data with $21\,$cm interferometry.

We will illustrate our ideas by focusing on cosmic voids, though much of what
we say could be applied to protoclusters as well.
Our goal in this paper will thus be the detection of voids, and the challenges
associated with this.  We assume that these candidate voids will be
appropriately analyzed or followed up for different science applications.
It is worth keeping in mind that, for certain applications, the intermediate
step of constructing an explicit void catalog and characterizing its purity
and completeness may not be necessary.  One might be able to construct
estimators of the quantities of interest directly from the visibilities.
We shall not consider such approaches in this paper.

The outline of the paper is as follows.
In \S\ref{sec:review} we establish our notation and provide some background
on interferometry, foregrounds for $21\,$cm experiments and matched filters.
Section \ref{sec:sims} describes the numerical simulations that we use to
test our matched filter and the profiles of voids in those simulations.
Our main results are given in \S\ref{sec:results} and we present our
conclusions in \S\ref{sec:conclusions}.
We relegate a number of technical details to a series of appendices.
In particular Appendix \ref{app:noise} discusses instrument noise for an
interferometer in the cosmological context,
Appendix \ref{app:transit} describes the formalism of transiting telescopes
(including cylinder telescopes) and the flat-sky limit, and
Appendix \ref{app:HI} discusses the manner in which neutral hydrogen might
be expected to trace the matter field at intermediate redshift.

\section{Background and review}
\label{sec:review}

In this section we provide some background information, to set notation and
provide an easy reference for our later derivations.

\begin{table}
\begin{center}
\begin{tabular}{cccccc}
  \hline
  $z$   & $\nu$  & $\chi$& $\mu$ &  $k/D_{10}$ &  $|d\chi/d\nu|$ \\
        & [MHz]  & [$h^{-1}$ Mpc] & & [$h {\rm Mpc}^{-1}]$& [$h^{-1}$ Mpc MHz$^{-1}$]\\ \hline
 0.50   &  947   & 1322  & 0.359 & 0.1509  & 3.630 \\
 0.75   &  811   & 1854  & 0.473 & 0.0922  & 4.256 \\
 1.00   &  710   & 2313  & 0.562 & 0.0647  & 4.796 \\
 1.25   &  631   & 2709  & 0.632 & 0.0491  & 5.267 \\
 1.50   &  568   & 3055  & 0.687 & 0.0392  & 5.685 \\
 1.75   &  516   & 3358  & 0.732 & 0.0324  & 6.061 \\
 2.00   &  473   & 3626  & 0.767 & 0.0275  & 6.405 \\
 2.25   &  437   & 3865  & 0.796 & 0.0238  & 6.724 \\
 2.50   &  406   & 4079  & 0.820 & 0.0210  & 7.023 \\
 \hline
\end{tabular}
\caption{Useful quantities and conversion factors as a function of redshift, $z$,
  assuming a flat $\Lambda$CDM model with $\Omega_m=0.3$. These are
  (a) the observing frequency, $\nu$, in MHz for $21\,$cm radiation emitted at $z$,
  (b) the comoving distance to $z$, $\chi$, in $h^{-1}$Mpc,
  (c) the foreground wedge angle [Eq.~\ref{eqn:wedge_angle}]
  (d) the $k$ mode which a $10\,$m baseline maps to at redshift $z$, and
  (e) the differential conversion from frequency (in MHz) to comoving distance (in $h^{-1}$Mpc).}
\label{tab:conversions}
\end{center}
\end{table}

\subsection{Visibilities}

In an interferometer the fundamental datum is the correlation between two
feeds (or antennae), known as a visibility.  For an intensity measurement
the visibility is \citep{TMS}
\begin{equation}
  V_{ij}\propto \int d^2n\ A^2(\vechat{n})
  T(\vechat{n})\ e^{2\pi i\vechat{n}\cdot\vec{u}_{ij}}
\end{equation}
where $T(\vechat{n})$ is the brightness temperature in the sky direction
$\vechat{n}$, $A(\vechat{n})$ is the primary beam (assumed the same for all feeds) and
$\vec{u}_{ij}$ is the difference in position vectors of the $i^{\rm th}$
and $j^{\rm th}$ feeds in units of the observing wavelength.
It is common to normalize the visibilities so that they return brightness
temperature.  We convert from brightness temperature to cosmological
overdensity throughout, so we omit the exact normalization here.
We will work in visibility space, since this is the natural space for the
interferometer and has the simplest noise properties.
Some useful conversions between common quantities are given in
Table \ref{tab:conversions}.

Visibilities are measured over a range of frequencies, and we shall follow
the common procedure in $21\,$cm studies of Fourier transforming in the
frequency direction to obtain a data cube in 3D Fourier space, $\vec{k}$.
The conversion from frequency to distance (and hence Fourier mode) is
\begin{equation}
  \left|\frac{d\chi}{d\nu}\right| = \frac{c}{H(z)}\ \frac{(1+z)^2}{\nu_0}
\end{equation}
with $\nu_0=1420\,$MHz.

For small sky areas, the visibility thus measures the Fourier transform of
the sky signal, apodized by the primary beam.
Approximating the sky as flat and assuming 
the signal of interest, $\tau$, is azimuthally symmetric
(see Appendix \ref{app:transit})
\begin{equation}
  \tau(k) = 2\pi\int\tilde{\omega}\,d\tilde{\omega}
  \ J_0(k\,\tilde{\omega})\,\tau(\tilde{\omega})
\end{equation}
where $\tilde{\omega}$ is an angular, radial coordinate and
\begin{equation}
  V_{ij} \propto \left[\tau\star B\right](2\pi u_{ij})
\end{equation}
with the $\star$ representing a convolution and $\tau$ and $B$ being the
Fourier transforms of $\tau(\vechat{n})$ and $A^2(\vechat{n})$ respectively.
The surveys of interest to us here will cover large sky areas.  The
(very correlated) visibilities from the different pointings can be combined
to produce higher resolution in the $u-v$ plane (a process known as
mosaicking), entirely analogously to the manner in which the many slits in
a diffraction grating sharpen the transmitted lines
(e.g.~\citealt{TMS}, or see the discussion in \citealt{Whi99} for the
cosmological context).
In such a case, the effective $B$ is determined by the survey area rather than
the primary beam (analogous to a survey window in a galaxy survey) and will be
very small.  Combined with the fact that our signals will be very smooth in
$u-v$ this allows us to neglect $B$ to simplify our presentation.
Reinstating it does not change any of our conclusions.

\subsection{$21\,$cm interferometers}

We shall start by considering an interferometer consisting of an array of
dishes (the interesting case of transiting, cylinder telescopes presents
only technical modifications and is described in Appendix \ref{app:transit}).
As a concrete example, we use the HIRAX experiment \citep{HIRAX}.
HIRAX will use 1024 $6\,$m parabolic dishes in a compact grid covering the
frequency range $400<\nu<800\,$MHz (i.e.~$0.8<z<2.5$ for $21\,$cm radiation).
HIRAX is a transit telescope: all dishes will be pointed at the meridian
with a given declination, and the sky will rotate overhead in a constant
drift-scan.  Each declination pointing will give access to a $6^\circ$
wide stripe of the sky and the complete survey will cover $15,000$
square degrees.

\begin{figure}
\begin{center}
\resizebox{\columnwidth}{!}{\includegraphics{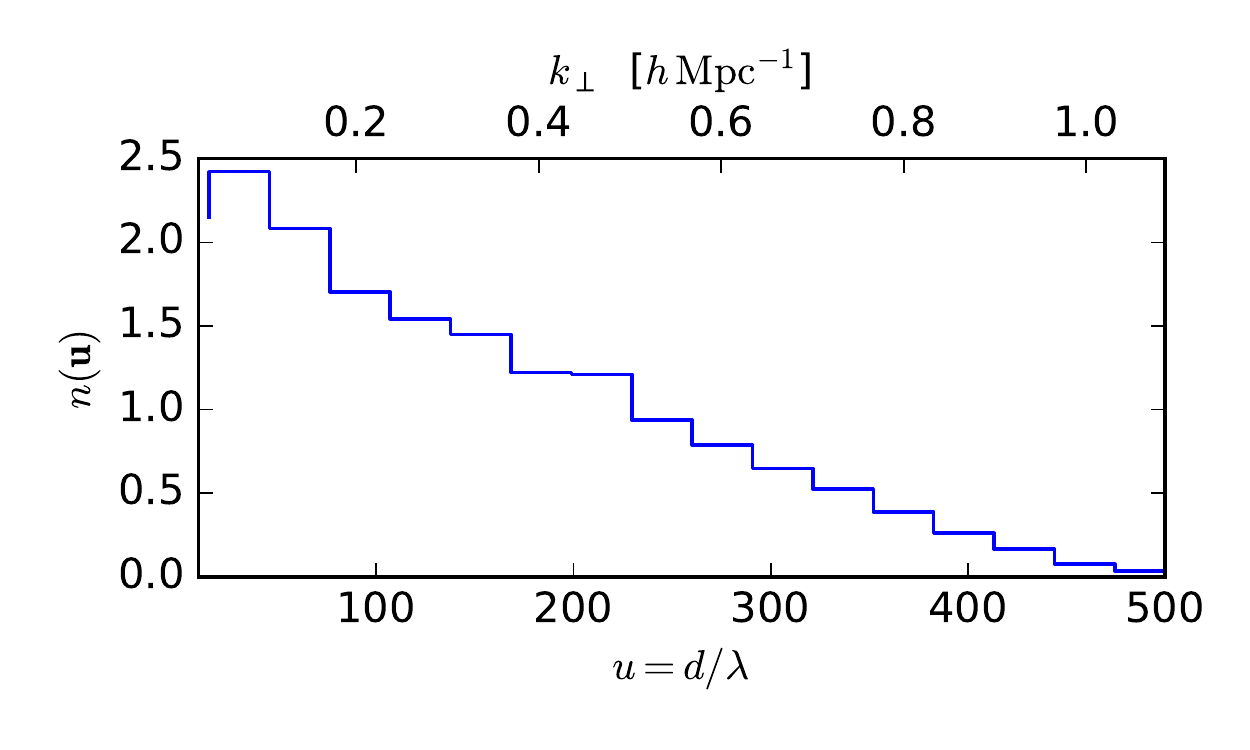}}
\end{center}
\caption{The circularly averaged baseline distribution for a HIRAX-like
experiment.  The lower axis shows the length baseline separation in units
of the wavelength for $\lambda=42\,$cm (i.e.~$z=1$) while the upper axis
converts to $k_\perp$ in $h\,{\rm Mpc}^{-1}$. The distribution is normalized
to integrate to the number of antenna pairs.}
\label{fig:baselines}
\end{figure}

Fig.~\ref{fig:baselines} plots the circularly averaged distribution of
baselines at $z=1$, given our assumptions for HIRAX.
The $x$-axis is the baseline separation in units of the wavelength,
$|\vec{u}|$, while the $y$-axis is the number density of baselines
per $d^2u$, conventionally normalized to integrate to the number of
antenna pairs.
The upper $x$-axis shows what $k$ modes these baselines map to.
Recalling that the noise scales as the inverse of the baseline density
(see Appendix \ref{app:noise}),
we see that a HIRAX-like experiment is sensitive to a broad range of $k$
scales well suited to the detection of voids and protoclusters.

\subsection{Noise for $21\,$cm experiments}
\label{sec:noise}

The major difficulty facing upcoming $21\,$cm experiments is astrophysical
foregrounds \citep[e.g.][]{Fur06,Sha14,Sha15,Pob15,SeoHir16}.
Foregrounds have been extensively studied in the context of (high $z$)
epoch of reionization studies, i.e.~at lower frequencies than of direct
interest here.  However the amplitudes of the signal and foreground scale
in a roughly similar manner with frequency, so many of the lessons hold in
our case \citep[see][for a recent discussion]{Pob15}.
Since the main (Galactic) foregrounds are relatively smooth in frequency,
their removal impacts primarily the slowly varying modes along the line of
sight, i.e.~the low $k_\parallel$ modes.  However since the foreground are
very bright (compared to the signal) and no instrument can be characterized
perfectly, there is also some leaking of foreground power into other parts
of the $k_\perp-k_\parallel$ plane.

The precise range of scales accessible to $21\,$cm experiments
after foreground removal is currently a source of debate.  We do not attempt
to model foreground subtraction explicitly, but take into account its effects by
restricting the range of the $\vec{k}_\perp - k_{\parallel}$ plane we use.

There are two regions of this plane we could lose to foreground removal.
The first is low $k_\parallel$ modes, i.e.~modes close to transverse to the
line-of-sight.
This boundary is slightly fuzzy and not well known.
For instance, \citet{Sha14,Sha15} claim that foreground removal leaves modes
with $k_\parallel> 0.02\,h{\rm Mpc}^{-1}$ available for cosmological use,
while \citet{Pob15} claims $k_\parallel<0.1\,h{\rm Mpc}^{-1}$ modes are
unusable.  We shall consider the impact of a $k_\parallel$ cut within this
range and we will see that our ability to find voids is quite sensitive to
this cut.

In addition to low $k_\parallel$, non-idealities in the instrument lead to
leakage of foreground information into higher $k_\parallel-k_\perp$ modes.
This is usually phrased in terms of a foreground ``wedge''
\citep[for recent discussions see][and references therein]{Pob15,Sha15,
SeoHir16,Coh16}.
The wedge does not form a hard boundary, but delineates a region where modes
far from the line-of-sight direction can become increasingly contaminated.
For a spatially flat Universe we can define the wedge geometrically as
\citep{SeoHir16,Coh16}
\begin{equation}
  \mathcal{R} 
  = \frac{\chi\,H}{c(1+z)} = \frac{E(z)}{1+z}\int_0^z\frac{dz'}{E(z')}
\end{equation}
where $E(z)=H(z)/H_0$ is the evolution parameter, and we assume we cannot
access the signal in modes with $|k_\parallel|/k_\perp<\mathcal{R}$ or
\begin{equation}
  \frac{|k_\parallel|}{k} < \mu_{\rm min}
  = \frac{\mathcal{R}}{\sqrt{1+\mathcal{R}^2}}
  \approx 0.6
\label{eqn:wedge_angle}
\end{equation}
with the last step being for $z=1$. It is worth emphasizing that this
foreground ``wedge'' does not represent a fundamental loss of information,
and may be mitigated with an improved model of the instrument (ideally the
wedge can be reduced by $\sin\Theta$ where $\Theta$ is the field of view;
\citealt{Liu16}).
We bracket these cases by considering cases with and without the foreground
wedge, and discuss the impact on our void finder.

Finally we must contend with shot-noise and receiver noise in the instrument.
\citet{CasVil16} argue that shot-noise is sub-dominant to receiver noise for
upcoming surveys, so we shall neglect it in what follows
\citep[see also][]{Coh16}.
To simplify our presentation we shall treat the receiver noise as
uncorrelated between visibilities and constant for all pairs of receivers.
The noise thus scales with the number of baselines that probe a particular
scale, and only an overall scaling is required.
If the noise is uncorrelated from frequency channel to frequency channel,
and only slowly varying with frequency, then the noise level is independent
of $k_\parallel$.
It is convenient to quote the thermal noise power in terms of the linear
theory power spectrum, $P_L$, in much the same way as galaxy surveys specify
their shot-noise by giving $\bar{n}P$ at some fiducial scale.
Since one of the design goals of all of these surveys is a measurement of the
baryon acoustic oscillation (BAO) scale,
we follow the standard practice and specify the receiver noise as a fraction
of $P_L$ at $k_{\perp,{\rm fid}}=0.2\,h\,{\rm Mpc}^{-1}$.
The surveys should achieve $P_L/P_{\rm noise}>1$ at $k_{\perp,{\rm fid}}$ and we
shall explore a range of values (see Appendix \ref{app:noise}).
Once $P_L/P_{\rm noise}>3$ the results become very insensitive to the
precise value.

\subsection{Matched filters}

A matched filter is a convenient means of finding a signal of known shape
in a noisy data set.  If we write the the data as an amplitude times a
template plus Gaussian noise ($d=A\tau+n$), the maximum-likelihood estimate
of $A$ and its scatter is given by
\begin{equation}
  \hat{A} = \frac{\tau N^{-1} d}{\tau N^{-1}\tau} \qquad , \qquad
  \sigma^{-2} = \tau N^{-1} \tau
\label{eq:matchedfilter}
\end{equation}
We take the ``noise'' covariance to include both instrument noise and
non-template cosmological signal and shall assume throughout that this
noise is diagonal in $k$-space.
The main feature of this expression is that areas
of the $k$-plane which are not sampled or are lost to foregrounds
receive zero weight ($N^{-1}=0$).

We find that our ability to isolate voids is very insensitive to the
exact profile chosen for $\tau$.  In fact even a top-hat profile produces
a highly pure and complete void catalog for low noise and good $k$-space
sampling.  Similarly the performance is not particularly sensitive to the
particular choice for $N(k)$, but rather to the larger questions of whether
there are significant regions of $k$-space where $N^{-1}=0$ or very uneven
sensitivity of the instrument due to the spacing of the feeds.

We shall use N-body simulations for our signal, and work with a periodic,
cubic box.  In such situations, given a 3D density field, $\delta(\vec{x})$,
and a template, $\tau(\vec{x})$, we can implement the flat-sky version of
the matched filter very efficiently using FFTs if the noise is diagonal in
$k$-space.
Recalling that a shift in configuration space amounts to multiplication by a
phase in Fourier space, the matched filter for a void centered at $\vec{a}$ is
\begin{equation}
  \tau N^{-1}d \to \sum_{\vec{k}} e^{i\vec{k}\cdot\vec{a}}
  \ \frac{\tau_0(\vec{k})\delta^\star(\vec{k})}{N(\vec{k})}
\end{equation}
where $\tau_0$ is the template for a void centered at the origin.
The sum is simply an (inverse) Fourier transform, so we can test for all
$\vec{a}$ at once.
A similar set of steps can be used for the denominator $\tau N^{-1} \tau$,
allowing a fast computation of $S/N$ for any position, $\vec{a}$.
Thus with forward Fourier transforms of the template and data and one
inverse transform we can compute the matched filter amplitude, $A$,
everywhere in space and hence its (volume weighted) distribution at
random locations and at the positions of voids.

\begin{figure*}
\begin{center}
\resizebox{6in}{!}{\includegraphics{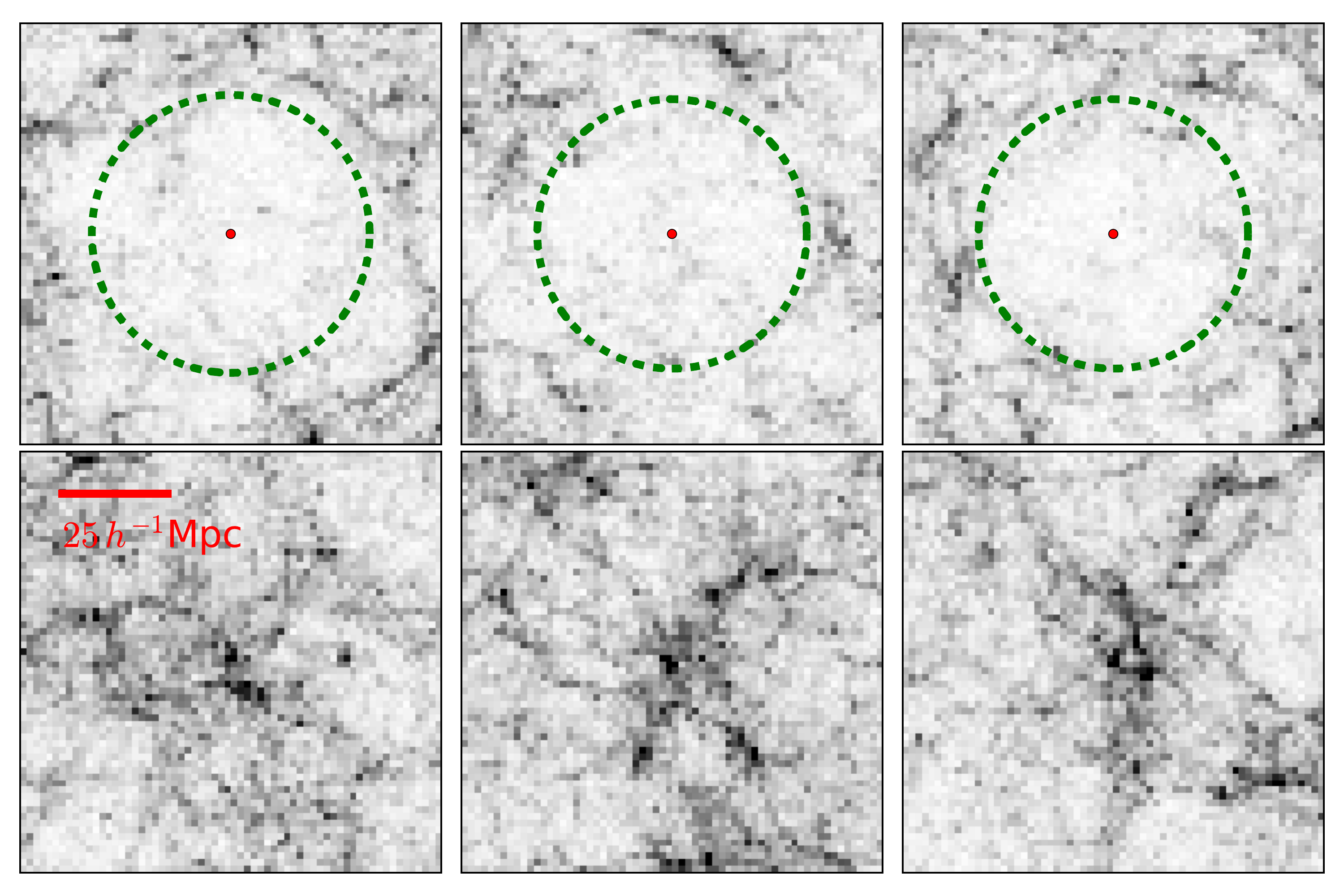}}
\end{center}
\caption{Slices through one of our N-body simulations at $z=1$.  Each
panel shows the projected, redshift-space density field, on an arcsinh scale
saturating at $10\,\bar{\rho}$, with a (line-of-sight) depth of
$\pm 20\,h^{-1}$Mpc and transverse dimensions $\pm 50\,h^{-1}$Mpc.
The panels in the top row are centered on voids, with the void center marked
with a dot and the radius with a dashed circle.  Those in the bottom row are
centered on (randomly selected) massive protoclusters.}
\label{fig:slice}
\end{figure*}

There is, in principle, no reason why the matched filter can't be modified to
remove spectrally smooth foregrounds at the same time as searching for voids
or protoclusters.  We choose not to implement this approach, preferring
instead to set $N^{-1}=0$ for modes which we deem unusable due to foregrounds.

\section{Simulations}
\label{sec:sims}

\subsection{N-body}

To illustrate our ideas we make use of several N-body simulations, each of
the $\Lambda$CDM family.  Specifically we use the $z=1$ outputs of 10
simulations run with the {\sl TreePM} code described in \citet{TreePM}.
This code has been extensively compared to other N-body codes in
\cite{Hei08}, and these simulations have been previously used (and described)
in \citet{Rei14} and \citet{Whi15}.  Each run utilized $2048^3$ particles in a
periodic box of side $1380\,h^{-1}$Mpc to model a cosmology with
$\Omega_m=0.292$, $h=0.69$ and $\sigma_8=0.82$.  This gives a particle
mass of $m_p=2.5\times 10^{10}\,h^{-1}M_\odot$.
Slices of the redshift-space density field at $z=1$ from one of the
simulations are shown in Fig.~\ref{fig:slice}, to illustrate the types of
structures we are searching for.  There are spatially coherent regions of
over- or under-density with scales of $\mathcal{O}(10\,{\rm Mpc})$
clearly visible in the figure.

Properly modeling the distribution of HI at $z=1-2$ is beyond the scope of
this paper.  Our simulations would need much higher resolution, to resolve
the halos likely to host neutral hydrogen at $z=1-2$, and the halo occupancy
is anyway highly uncertain \citep[see discussion in e.g.][]{CasVil16,See16}.
Instead we assume the HI is an unbiased tracer of the matter field, and
simply use the dark matter density.  In Appendix \ref{app:HI} we use a
halo model of HI in a higher resolution (but smaller volume) simulation
to show that this is a conservative approximation for the purposes of
establishing how well $21\,$cm experiments can find voids.

\subsection{Voids in the Simulations}

We define voids through a spherical underdensity algorithm (for a comparison
with other void finders see \citealt{Sta15b}, for a general comparison of
void finders see \citealt{Col08}).  The dark matter particles are binned onto
a regular, Cartesian grid of $1380^3$ points.  Around each density minimum
with $1+\delta<0.2$ we grow a sphere until the mean enclosed density is
$1+\bar{\delta}<0.4$.  Visually such an underdensity gives voids which match
expectations (see Fig.~\ref{fig:slice}).
The voids are then ordered by their radius $R_V$ and overlapping voids with smaller radii
are removed from the list.
As is the case for the large overdensities (protoclusters) these large
underdensities (voids) are very rare, necessitating surveys of large volumes.
The number density of redshift-space voids at $z=1$ is $10^{-5}$ $h^{-3}{\rm Mpc}^3$ for
$10<R_v<15\,h^{-1}$Mpc and $6\times 10^{-7}$ for $20<R_v<25\,h^{-1}$Mpc
and falls quickly with redshift.


The matched filter essentially performs a ``weighted convolution'' of the
density field with a profile, and thus requires some knowledge of the shape
of the object it is trying to `match'.  
While the performance of the filter is relatively insensitive to the
precise profile we use, we describe the choices we have
made based on the N-body simulations described above.

To begin we note that a void has an extent
$\mathcal{O}(10\,{\rm Mpc})$, and thus covers only a small region of sky ($< 1$ arcminute)
and a small portion of the frequency coverage of the telescope.
We are thus justified in treating the sky
as locally flat and the $\vec{k}_\perp$ coverage as approximately wavelength
independent \footnote{Recall that the conversion from $\vec{u}$ to $\vec{k}_{\perp}$ depends
on the frequency of the observation.}.  We expect the profile to have significant power at
$k\sim 0.1\,h\,{\rm Mpc}^{-1}$, well within the band of sensitivity of
$21\,$cm interferometers aimed at large-scale structure observations (see
Fig.~\ref{fig:baselines} before).

\begin{figure}
\begin{center}
\resizebox{\columnwidth}{!}{\includegraphics{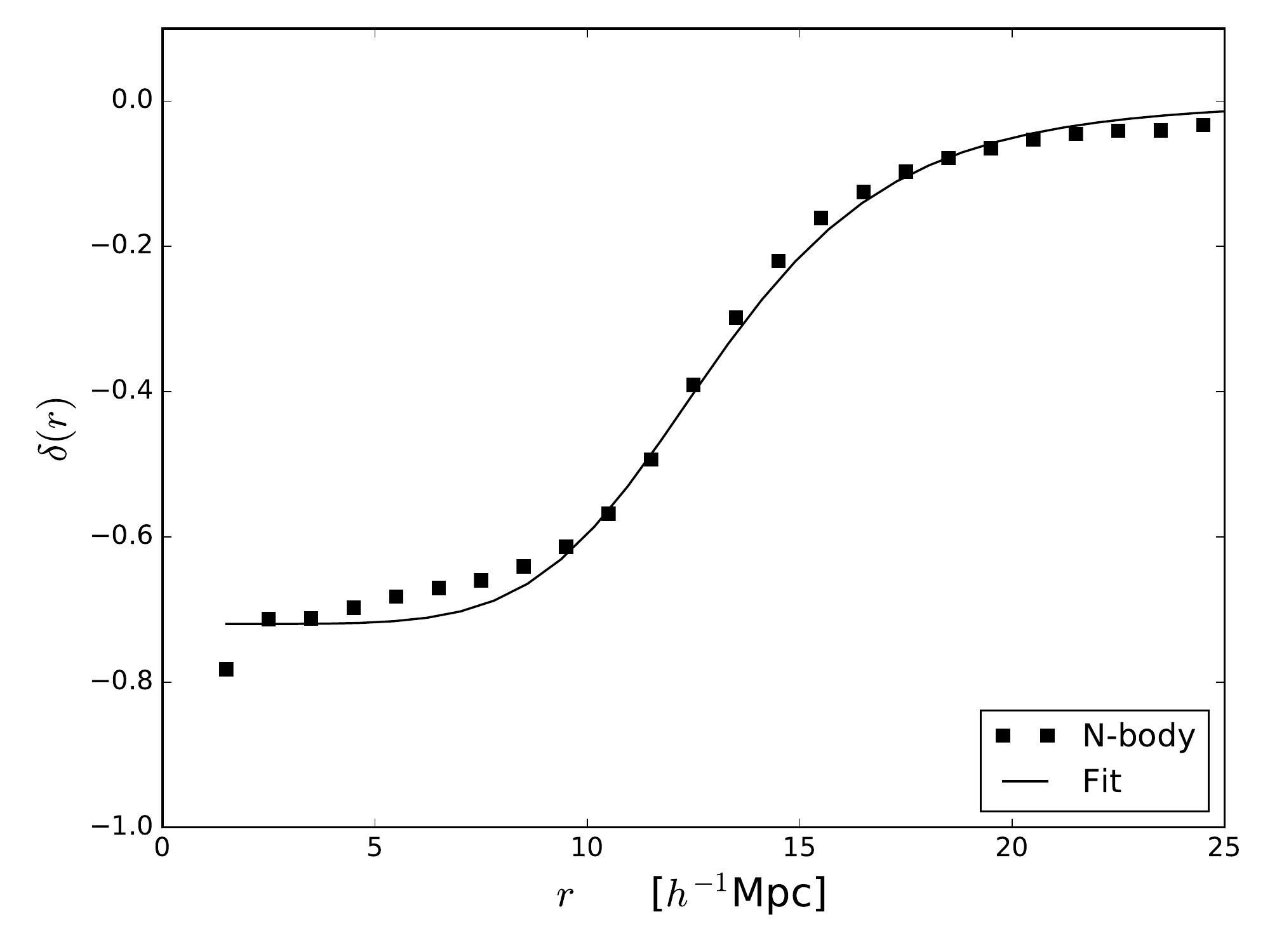}}
\caption{Real-space profile of voids in our simulation.
Squares show the average profile measured from the N-body simulations described in
the text for voids with radii $10<R_v<15\,h^{-1}$Mpc at $z=1$.
The solid line is the analytic fit of Eq.~\ref{eqn:our_profile} with
$r_s=12.5\,h^{-1}$Mpc.
}
\label{fig:profile1d}
\end{center}
\end{figure}

Fig.~\ref{fig:profile1d} shows the averaged real space profile of voids with
$10 < R_V < 15$ $h^{-1}$Mpc 
in our N-body simulations at $z=1$.
There are numerous analytic void profiles in the literature
\citep[e.g.][for recent examples]{Ham14,Haw17}.
Most of these do not fit our N-body results particularly well, which is most
likely due to the different choices of void finder employed.  In particular
our void profile approaches $0$ smoothly from below at large radius, i.e.~we
do not find a prominent ``compensation wall'' at the void edge.
This echoes the findings of \citet{Cai16}, who also found no compensation wall
for voids which are not part of a larger overdensity.



A simple, 2-parameter form which does provide a good fit to our N-body data is
\begin{equation}
  \delta = \frac{\delta_0}{1+(r/r_s)^6}
\label{eqn:our_profile}
\end{equation}
where $\delta_0$ and $r_s$ are the interior underdensity and void scale
radius respectively.  This is shown in Fig.~\ref{fig:profile1d} as the
solid line.  In Fourier space this profile becomes
\begin{eqnarray}
  \tau(k) &=& 4\pi\int r^2\,dr\ \delta(r)\,j_0(kr) \\
       &=& \frac{2\pi^2}{3\,\kappa}\delta_0\,r_s^3\,e^{-\kappa/2}\left[
           e^{-\kappa/2} + \sqrt{3}\sin y-\cos y\right]
\end{eqnarray}
where $\kappa=k\,r_s$ and $y=\sqrt{3}\,\kappa/2$. 
For large $k$ the profile is exponentially suppressed.
Since the profile is not compensated,
$\tau(k\ll 1)\simeq (2\pi^2/3)(1-\kappa^2/3+\cdots)\delta_0 r_s^3$ does not
go to zero as $k\to 0$.
This is clearly only an approximation, since on sufficiently large scales
the profile must go to zero due to mass conservation, but it does not seem
to adversely affect our filter. We remind the reader that it is this
Fourier space form that is input into the matched filter.

\begin{figure}
\begin{center}
\resizebox{\columnwidth}{!}{\includegraphics{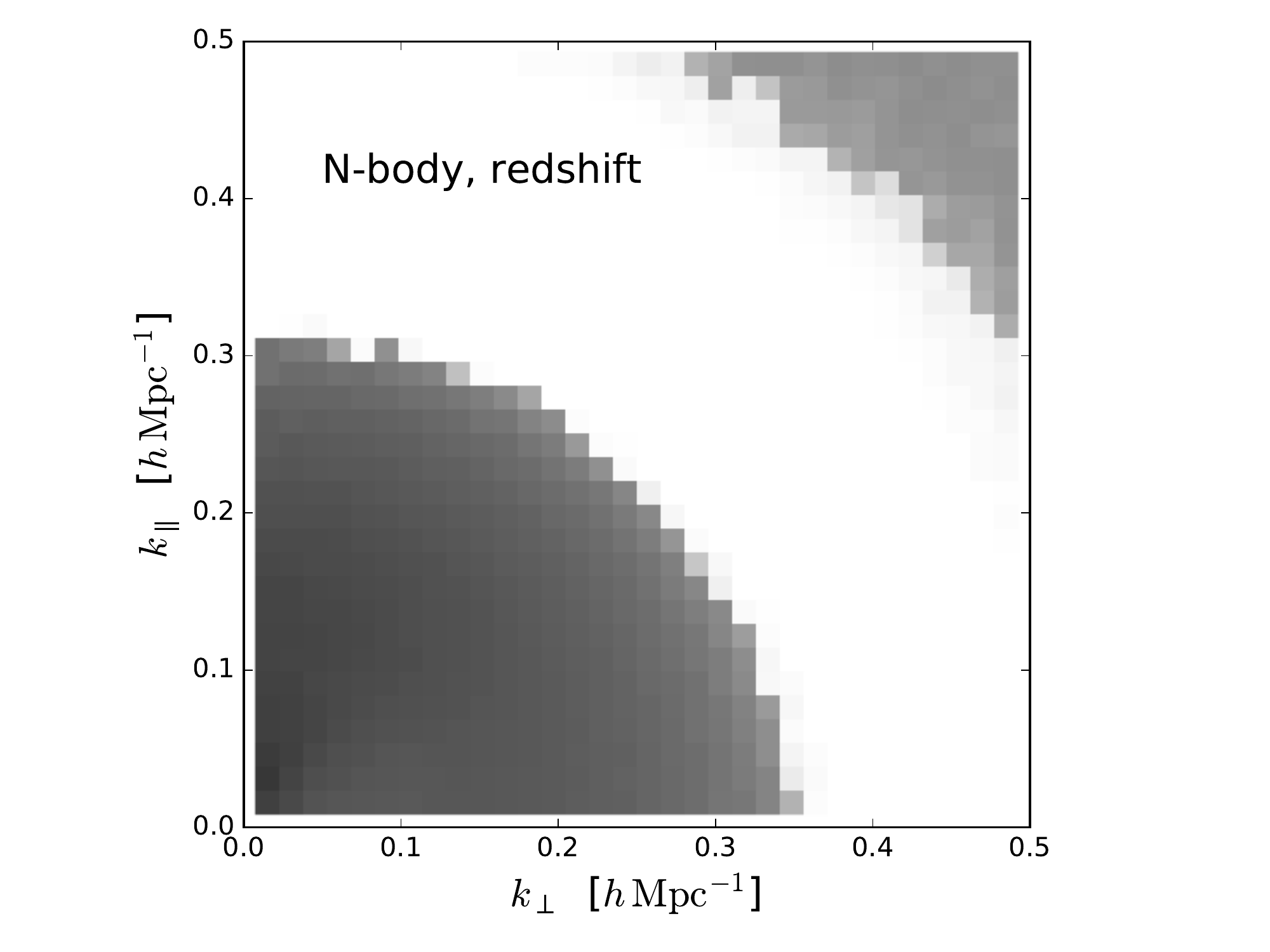}}
\caption{The (stacked) profile in Fourier space for voids
  with radii $10<R_v<15\,h^{-1}$Mpc from our N-body
simulations in redshift space at $z=1$.}
\label{fig:profile_compare}
\end{center}
\end{figure}

The above was all in real space.
An analytic model for a void in redshift space could simply use the linear
theory analysis of \citet{Kai87}.  A better alternative would be to make use
of the Gaussian streaming model \citep{ReiWhi11}.  \citet{Ham15} have shown
that this model works well if linear theory expressions for the mean
pairwise velocity and dispersion are computed from the assumed profile.
We have taken a simpler approach, using the simulations to measure the
anisotropy. Fig.~\ref{fig:profile_compare} shows the Fourier transform of the
same voids shown in Fig.~\ref{fig:profile1d}, except now in redshift space. 
Material which is outflowing causes the void to appear deeper
and wider in the line-of-sight direction \citep{Kai87}, enhancing the profile
along $k_\parallel$.  In principle, this makes the redshift-space profile less
sensitive to loss of modes in the ``wedge'' than would be anticipated from the
real-space profile (though the filter will tend to downweight the line-of-sight
modes more due to the enhanced cosmic clustering close to the line of sight).

However, for intermediate scales $k \sim 0.2 h{\rm Mpc}^{-1}$, the void
profiles are remarkably close to spherical, with only a very mild quadrupole. 
Given this small anisotropy, we shall continue to use a spherically symmetric
void profile even in redshift space. This choice was motivated purely for the
simplicity of the presentation and does not represent a limitation of the
method, and we expect these choices to be revisited in future work.

\section{Results}
\label{sec:results}

\subsection{Filter amplitude distributions}

\begin{figure*}
\begin{center}
\resizebox{6.7in}{!}{\includegraphics{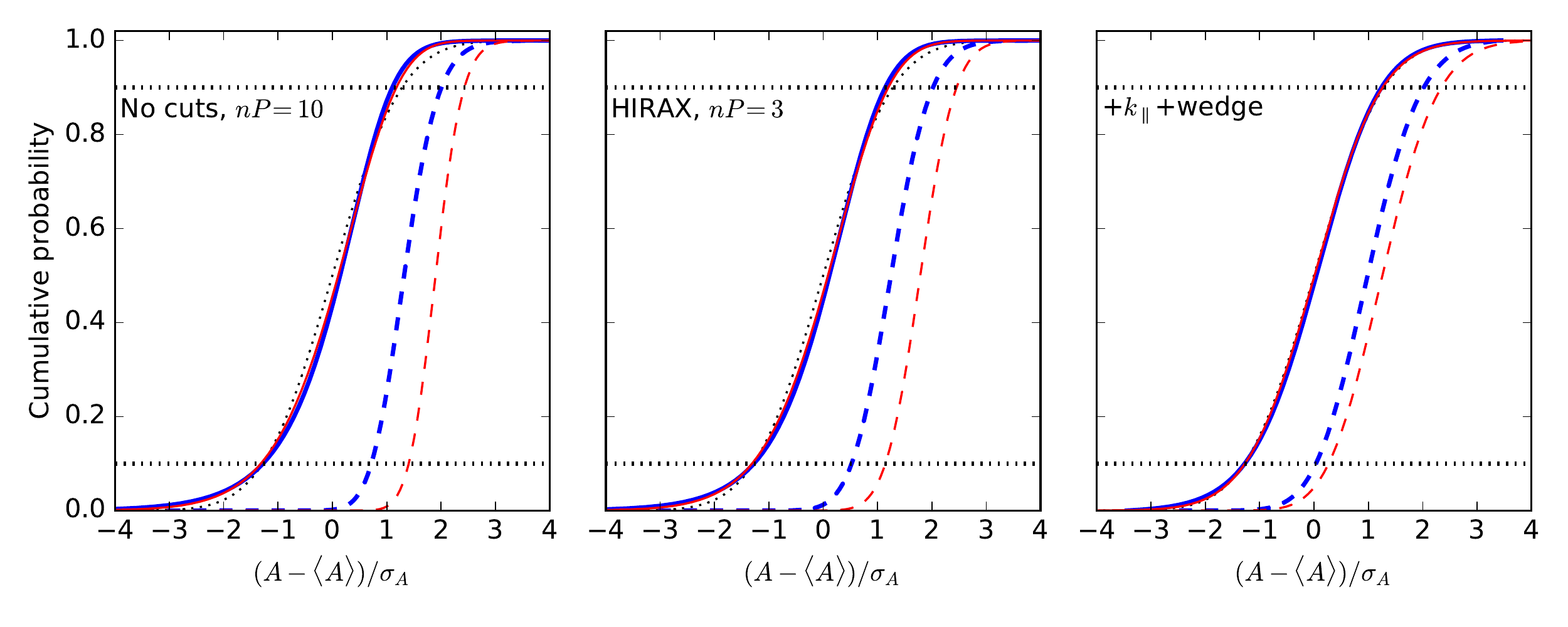}}
\caption{The (normalized) distribution of matched filter amplitude, $A$,
at the locations of voids (dashed) compared to the full (volume weighted)
distribution (solid) for voids of radius $10-15\,h^{-1}$Mpc (thick, blue)
and $20-25\,h^{-1}$Mpc (thin, red).  In each panel the dotted (black) line
shows a unit-variance Gaussian for reference.  The left panel shows void
recovery for perfect sampling of the $k_\perp$ plane with minimal noise
(we have quoted the noise as $\bar{n}P=10$ at
$k_{\rm fid}=0.2\,h\,{\rm Mpc}^{-1}$, in analogy with galaxy surveys but
recall $\bar{n}$ is in reality thermal noise as a function of $k_\perp$).
The middle panel shows a baseline distribution for a HIRAX-like telescope
with $\bar{n}P=3$ and the right panel shows the additional effects of
removing modes with $k_\parallel<0.05\,h\,{\rm Mpc}^{-1}$ and $\mu_k<0.56$.
The horizontal dotted lines in each panel mark the $10^{\rm th}$ and
$90^{\rm th}$ percentiles for reference.}
\label{fig:amp_dist}
\end{center}
\end{figure*}

We now turn to the performance of the matched filter.
Recall that we can evaluate the matched filter at an arbitrary
point - ideally positions centered on voids would have significantly
larger values of $\hat{A}$ that a randomly chosen point.
\footnote{Since our input void profile has a negative central
  underdensity, we expect voids to have positive values of $\hat{A}$.}
The left panel of Fig.~\ref{fig:amp_dist} plots the distribution of
$\hat{A}$ in the ideal
case of an effectively noiseless $\bar{n} P=10$ survey. The distribution
is close to Gaussian with a width of 0.86; this compares with the
analytically predicted value (Eq.~\ref{eq:matchedfilter}) of 0.90.
The Gaussianity of this distribution is easily understood by observing
that the matched filter simply smoothes the (configuration space) density
field with a kernel that is ${\cal O}(10)$ Mpc wide; on these scales, the
density field is very close to Gaussian. We do see evidence of non-Gaussianity
from collapsed objects in a slight skew towards negative values of ${\hat A}$.
Although the matched filter has the void radius as an input parameter,
we find that the shapes of the distributions (after scaling out the variance) are
very similar. We therefore simply standardize all our distributions by the
appropriate variance.

We can now compare the above distribution with the matched filter evaluated
at void centers.
We consider two sets of voids : $10 < R_V < 15 h^{-1} {\rm Mpc}$ and
$20 < R_V < 25 h^{-1} {\rm Mpc}$ and we set the filter radius to
$12.5\,h^{-1}$Mpc and $22.5\,h^{-1}$Mpc respectively.
We find that the distribution of $\hat{A}$ evaluated at the void centers
is clearly separated from the full distribution of the matched filter.
Approximately $80\%$ of the smaller voids are detected at $> 1 \sigma$ from
zero while $\sim 90\%$ of the larger voids are detected at $> 1.5 \sigma$.
It is also worth noting that our reference distribution includes points that
are in voids.
Indeed $\sim 8\%$ of the simulation volume is contained in voids larger than
$10 h^{-1} {\rm Mpc}$, which would correspond to a threshold choice of
$\sim 1\sigma$.
Note that this is somewhat different from the Gaussian expectation of
$\sim 1.5 \sigma$; this difference can be traced to non-Gaussianity in
the tails of the distribution of $\hat{A}$.

We now consider how survey non-idealities impact the efficiency of the
matched filter.
There are two aspects relevant to the $21\,$cm interferometer case.
The first is that the instrument only samples particular $k$-modes and
that this sampling is modulated by the number of baselines in the
interferometer.
The second is that, as discussed in Sec.~\ref{sec:noise}, astrophysical
foregrounds and instrumental imperfections can contaminate both low
$k_{||}$ modes and the so-called ``wedge'', further restricting the
accessible $k$-space.
The impact of these are summarized in the middle and right panels of
Fig.~\ref{fig:amp_dist}.
The relatively wide and dense coverage in $k$-space of our HIRAX-like
survey implies that the filter's performance does not degrade significantly
compared to the ideal case.
Removing modes contaminated by foregrounds has a more significant effect.
While we still see a separation between voids and randomly chosen points,
only $50\%$ of the voids are now above the thresholds discussed above.

While the detailed performance of the void finder will depend on the
details of the interferometer, the principal conclusion of the above
discussion is that for the designs that are being considered voids are
relatively easily detected in the absence of foregrounds but the loss of
low $k_\parallel$ modes is a serious matter and some foreground mitigation
strategy is necessary.
Fig.~\ref{fig:amp_dist_chime} shows similar performance plots for a
idealization of the CHIME experiment (see Appendix~\ref{app:transit}
for details).
As with our HIRAX example, we find a clear separation between the
distribution of voids and random points with similar recovered fractions
of voids for the cases without any foregrounds, and a loss of separation
when foregrounds become important.

As with all matched filter applications, there are a number of input choices. The choice of the
void profile is the most notable example in this case. We experimented with different choices
of void shapes and sizes and find that the results above are quite robust. A different complication
arises from the fact that our void profiles are estimated from the dark matter. Appendix~\ref{app:HI}
explores the shapes of voids with a more realistic modeling of the 21cm density field. We find that
shapes of the voids here are very similar (and possibly more pronounced) to those in the dark matter.
We therefore expect our results to be qualitatively unchanged with more realistic modeling of the
21cm field.

Another choice in our matched filter is the power spectrum used in the
noise covariance matrix to account for large-scale structure noise.
While different choices here change the exact width of the distribution
of $\hat{A}$, it does not change our basic result that voids are detected
with very high significance in the absence of foregrounds.

\begin{figure}
\begin{center}
\resizebox{3.3in}{!}{\includegraphics{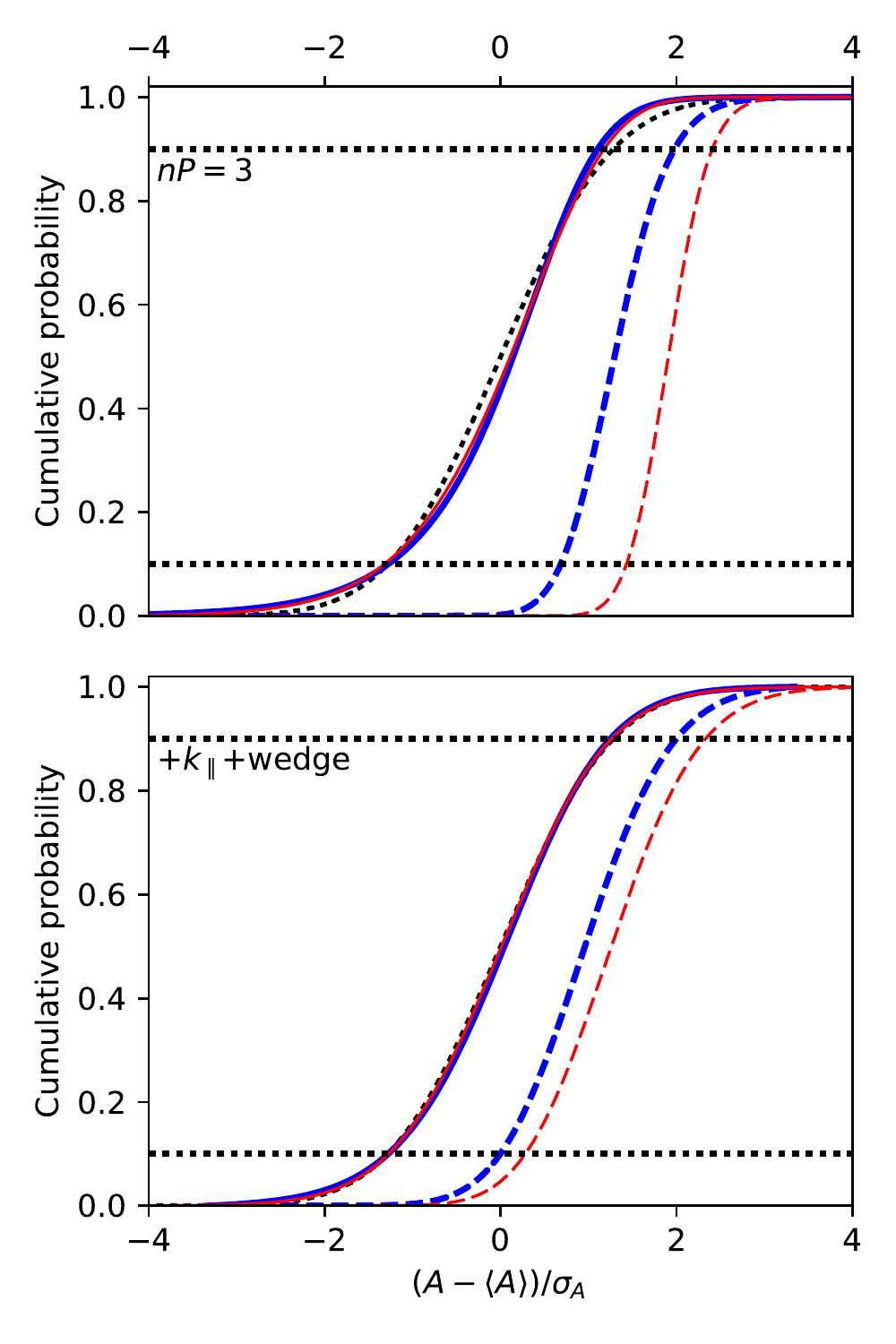}}
\caption{As in Fig.~\ref{fig:amp_dist} but for an idealization of the CHIME
telescope.  The top panel shows the impact of the CHIME baseline distribution
with a noise level appropriate for a BAO detection, while the lower panel
shows the additional impact of removing modes with
$k_\parallel<0.05\,h\,{\rm Mpc}^{-1}$ and $\mu_k<0.56$.}
\label{fig:amp_dist_chime}
\end{center}
\end{figure}

\subsection{An Example Application : A Void Catalog}

\begin{figure}
\begin{center}
\resizebox{\columnwidth}{!}{\includegraphics{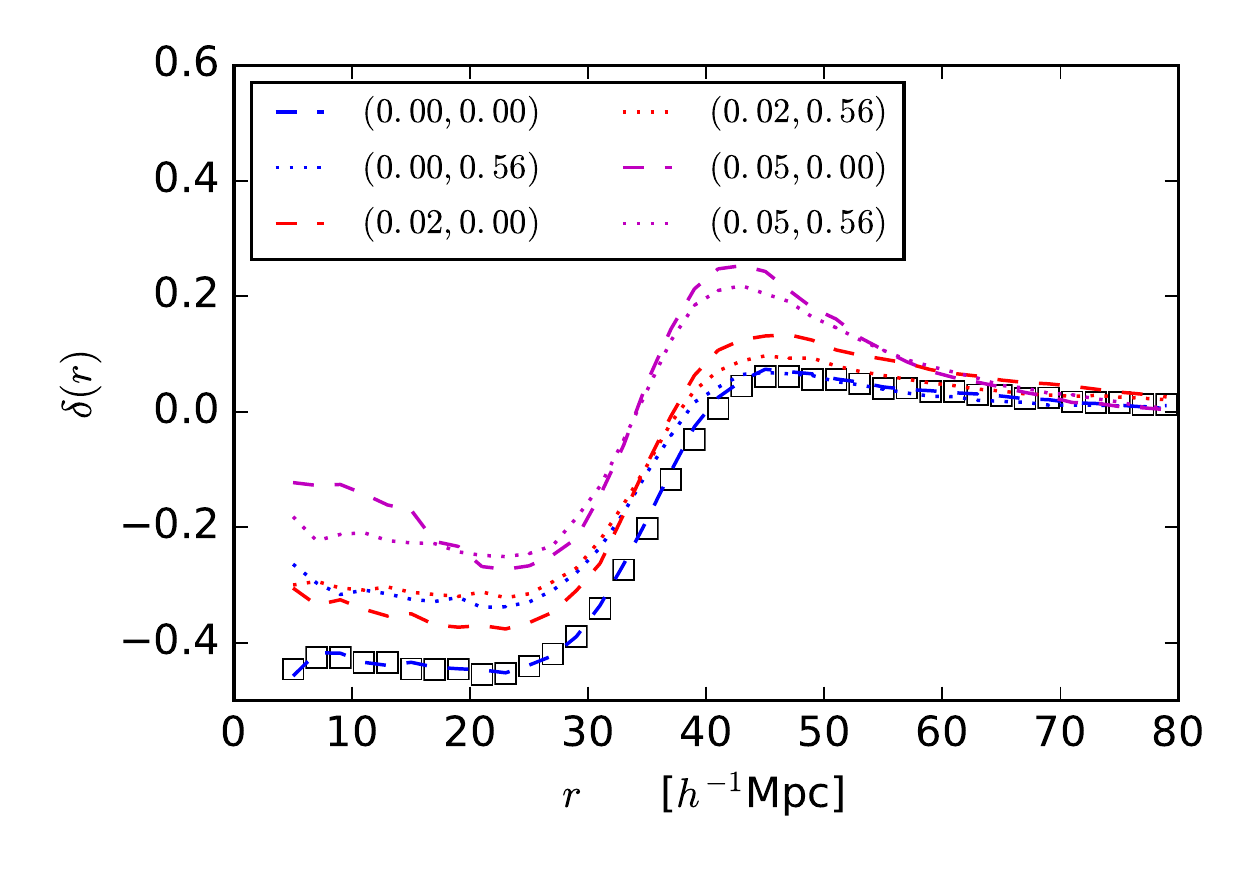}}
\caption{Redshift-space density profile stacked around the 1000 voids in our
``candidate catalog'', as described in the text.
Black squares show the stacked profile assuming perfect $u-v$ coverage and
no noise.  Clearly, in the absence of foregrounds, our candidates correspond
to large, coherent underdensities.
Compared to Fig.~\ref{fig:profile1d}, the shallower profile at small radius is
due to the miscentering described in the text.  The lines are all for the
HIRAX $u-v$ coverage with $\bar{n}P=3$, and show the impact of losing modes
at low $k_\parallel$ and in the wedge.  The legend gives the cuts as
$k_{\parallel,{\rm min}}$ and $\mu$ pairs.}
\label{fig:profile_Apeak}
\end{center}
\end{figure}

As an example application, we discuss how to use such a matched
filter to construct a void catalog. Our intention here is not to
attempt to quantify (or optimize) the purity and completeness
of such an algorithm, since this will be data and instrument specific 
and so much depends upon the manner in which foregrounds are subtracted.
Instead, we outline the steps of a possible algorithm and perform some
simple calculations with it, and defer detailed discussions to
future work.

For this demonstration, we choose a single simulation box from our
suite of ten simulations. We run the
matched filter on this box with the void radius $R_V$ varying from
$33.3\,h^{-1}$Mpc to $20\,h^{-1}$Mpc in 10\% steps.  We keep a list of all
points where the matched filter amplitude, $A$, exceeds $2\,\sigma$.
Starting from the largest void(s) and working down in radius we eliminate
any voids which overlap.  If two overlapping voids have the same radius the
one with the smallest $A$ is removed.  The result of this procedure is our
``void catalog''.

In a single $1380\,h^{-1}$Mpc box the largest 1,000 voids have radius above
about $20\,h^{-1}$Mpc.
With full $u-v$ coverage and low noise we find that all but 1 of the 32
largest ``true'' voids contain a match in our catalog within $0.75\,R_V$
and these matches are all in the upper $5^{\rm th}$ percentile of the
$A$ distribution.
Just under half of them (14 of the 32 voids)
show significant ($>R_V/3$) mis-centering, i.e. the detected void center is
$> R_V/3$ away from the center of the closest true void.
For the $u-v$ coverage of our HIRAX-like experiment, and $\bar{n}P=3$,
four of the 32 largest ``true'' voids do not have a match within $0.75\,R_V$
and again all are highly significant.
The situation changes dramatically as we include a $k_{\parallel,{\rm min}}$
and $\mu$ cut.  For  $k_{\parallel,{\rm min}}=0.05\,h\,{\rm Mpc}^{-1}$ and
$\mu>0.56$ we find only 5 of the top 32 voids in our catalog, though these
voids are in the extreme tails of the $A$ distribution.  Most of this effect
is driven by the $k_\parallel$ cut.
If we relax the cut to $0.02\,h\,{\rm Mpc}^{-1}$ then we recover 10 of the
32 largest voids and for a cut of $0.01\,h\,{\rm Mpc}^{-1}$ we recover 20
of them.

We can recast the results of this and the previous section into the more
traditional forms of the completeness and purity of the sample. In the absence
of foregrounds, our detected void catalog is both pure (only $\sim 10\%$ of
detected voids do not correspond to true voids) and complete ($> 90\%$ of true
voids are detected at better than $1.5\sigma$) for large ($\sim 20\,h^{-1}$Mpc)
voids. However, both of these numbers are sensitive to foregrounds. For our most
conservative case of foregrounds contaminating all modes with
$k_{||} < 0.05 h {\rm Mpc}^{-1}$ and $\mu < 0.56$, the majority of the most
prominent detections do not correspond to true underlying voids and only
$\sim 50\%$ of true voids are detected at high significance.

It is possible that some of the low $k_\parallel$ information lost to the
interferometer by foregrounds could be replaced by another experiment.
As an example, modern photometric surveys can achieve high photometric redshift
precision for certain types of galaxies, and thus can map the low $k_\parallel$
modes of the 3D density field.  In fact, such surveys have been used to
search for voids \citep{San17}.
Including the photometric survey in our matched filter presents no problem
in principle --- one simply augments the data vector and includes a model for
the void in configuration space --- but could be difficult in practice.
Assuming the combination recovers all of the $k_\parallel$ range, we recover
our no-foreground forecasts.
If there is a gap in coverage the results are adversely affected.
To take a pessimistic example: if we lose modes
$0.02<k_\parallel<0.05\,h\,{\rm Mpc}^{-1}$
we are able to recover 12 of our top 32 voids.  For
$0.03<k_\parallel<0.05\,h\,{\rm Mpc}^{-1}$
it is half of our top 32 voids.

These lost $k_\parallel$ modes potentially could be reconstructed from
higher-point information in the 21cm field itself \citep{ZhuTidal16}.
There is considerable interest in developing these reconstruction schemes
for 21cm surveys to enable cross-correlations with photometric surveys or
CMB lensing maps. 
Initial results \citep{ZhuTidal16} suggest that modes $k_\parallel < 0.01 h {\rm Mpc}^{-1}$ and
$k_\perp < 0.05 h {\rm Mpc}^{-1}$ could be recovered. 
As with the example above, the efficiency of the void finder will depend
on the details of the performance of these reconstructions.

\begin{figure}
\begin{center}
\resizebox{\columnwidth}{!}{\includegraphics{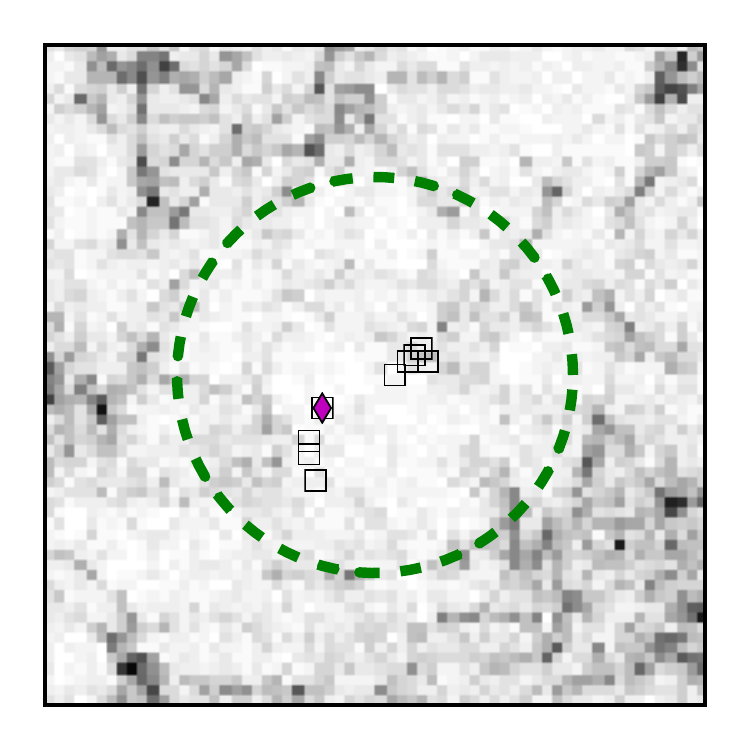}}
\caption{An example of a detected void in our catalog, for the case
  of no foregrounds. The image shows the matter distribution centered
  around a $R_V\simeq 30\,h^{-1}$Mpc void (bounded by the dashed line).
  The slice is $100\,h^{-1}$Mpc wide and $20\,h^{-1}$Mpc thick. The
  squares show significant matched filter detections, with the filled,
  magenta, diamond being the most significant detection (as determined by
  our pruning algorithm). We see that, while the void center is detected
  by the matched filter, it happens not to be the most significant detection,
  resulting in a mis-centered void.}
\label{fig:slice_z100_example}
\end{center}
\end{figure}

We can visualize this information in another way.
Fig.~\ref{fig:profile_Apeak} shows the stacked matter profile around our
top 1,000 void candidates for various choices of $k_{\parallel,{\rm min}}$
and $\mu_{\rm min}$.  With full $u-v$ coverage there is a clear, coherent
underdensity at the locations of the void candidates.  The shallower inner
profile in Fig.~\ref{fig:profile_Apeak} when compared to
Fig.~\ref{fig:profile1d} arises due to mis-centering.
While some voids were well centered, a significant fraction had offsets.
After visually inspecting these voids, we find that, in most cases, the true
void center was a significant detection in the matched filter, but happened not
to be the most significant detection and was removed by our relatively
simple pruning algorithm.
Fig.~\ref{fig:slice_z100_example} shows an example of such a case.
Both the fraction of found voids and the degree of mis-centering get worse
when modes are lost to foregrounds, as the other curves in
Fig.~\ref{fig:profile_Apeak} show.
For our pessimistic scenario of $k_\parallel>0.05\,h\,{\rm Mpc}^{-1}$
and $\mu>0.56$ there is barely any underdensity detected at all.
This suggests that doing science with voids selected via $21\,$cm experiments
will be difficult unless the foregrounds can be brought under control.
We however note that the algorithm used to construct our void catalog is
relatively simple; e.g.~a more robust algorithm might use multi-scale
information to get more robust measurements and there is significant potential
for complementarity between optical imaging surveys and $21\,$cm measurements. 

\section{Conclusions}
\label{sec:conclusions}

Recent advances in technology have made it feasible to study the $21\,$cm
emission from objects at cosmological distances.  A new generation of
telescopes is being designed and built which aim to survey enormous volumes
of the Universe with modest resolution at redshifts $z\simeq 1-2$.
A primary focus of these facilities is the measurement of the power spectrum
of large-scale structure, as traced by neutral hydrogen, which will hopefully
improve constraints on our cosmological model.
While these instruments do not have sufficient angular resolution to resolve
the emission from individual objects, we point out that they should be able
to make catalogs of the largest members of the cosmic web -- protoclusters and
voids -- if they are able to control foregrounds sufficiently.

We have considered instruments which measure the sky interferometrically,
which means they naturally operate in Fourier space.  The finite sampling
of the Fourier plane, and the loss of sensitivity in some modes due to
foregrounds, make it difficult to generate a real-space, 3D map from the
data and hence to search for exotica whose properties are not known in
advance.  However, our understanding of the cosmic web allows us to specify
in advance what sort of objects we are interested in finding and searches
for objects of known shape do not need to go through the map-making step:
a matched filter provides a natural method for finding such objects.
The matched filter formalism also allows us to mix multiple data sets, each
of which is provided in its own domain.

The cosmic web contains voids on a variety of scales, and voids which touch
or merge.  We have only studied the simplest matched filter.  The algorithm
can be modified to iteratively add voids to an existing catalog, always
adding the void which leads to the largest increase in the likelihood given
the already-found voids \citep[see e.g.][]{Koc03,Don08}.
This involves a scan over void (or protocluster) sizes, and increases the
complexity of the algorithm.
A multiprobe approach could use deep, optical imaging data in conjunction with
$21\,$cm data in much the same way as multifrequency information is sometimes
used for cluster finding \citep[e.g.][]{MelBarDel06,Ryk14}.
As our main aim was to assess the feasibility of void detection with $21\,$cm
surveys, we defer further consideration of such a process to future work.

Throughout we have focused our discussion on voids as exemplars of large
structures in the cosmic web.  Of course, the matched filter algorithm is
more general and the huge volume and sensitivity of upcoming experiments
can be used to search for a number of exotic objects.  At the other end of
the density distribution from voids are the large, coherent overdensities
associated with protoclusters.

Despite keen interest in the community in how clusters form and evolve,
and years of observational and numerical efforts, the study of early cluster
formation (at high $z$) remains observationally limited.
Protoclusters are rare, present only modest overdensities and lack many
of the features used to discover clusters (e.g.~a hot ICM or a red sequence).
Observations of protoclusters at high $z$ would provide important clues into
cluster assembly and the processes of galaxy formation \citep{Ove16}.
Given the diversity of protoclusters, having large samples with well
understood selection is important.  Like voids, protoclusters form large
coherent structures amenable to discovery in upcoming $21\,$cm
experiments.
Assuming a mean interior density of 200 times the background, the linear size
of the mean-density region from which material accretes into a present day
cluster is several (comoving) Mpc.  The progenitors of large clusters should
thus be identifiable in relatively low resolution maps that can cover large
volumes \citep[see e.g.][and Fig.~\ref{fig:slice}]{Ove16}.

Slices through the density field in one of our simulations are shown in
Fig.~\ref{fig:slice} where the large extended mass profile of the
protoclusters is evident.  In fact, the most massive clusters in the mature
Universe form not from the most overdense regions at high $z$ but from
large, possibly only moderately overdense regions such as shown in
Fig.~\ref{fig:slice} \citep{Ove16}.
While we do not show it here, the typical protocluster covers a larger
volume at $z\simeq 2$, rendering it potentially easier to see while still
being well within the redshift reach of HIRAX or CHIME.

The abundance of such protoclusters is identical to the abundance of the
clusters at $z=0$: for a mass threshold of $3\times 10^{14}\,h^{-1}M_\odot$
it is $4\times 10^{-6}\,h^3{\rm Mpc}^{-3}$.
This emphasizes the need for a survey to cover a large volume in order to
properly sample the heterogeneous population of protoclusters.
As an example, if it covered $15,000{\rm deg}^2$ between $z=1$ and 2 HIRAX
would survey $50(h^{-1}{\rm Gpc})^3$ encompassing $\sim 200,000$
protoclusters\footnote{Almost by definition the number density of protoclusters
is redshift independent.}.  CHIME is anticipated to cover a similar volume in
the northern hemisphere.
In some models the star formation associated with halos in protoclusters
makes up a significant fraction of the ionizing photon budget for
reionization \citep{Chi17} at $z\simeq 6-7$.  If foregrounds could be
controlled, using interferometers designed for studying reionization to
search for protoclusters could provide an interesting synergy.

\vspace{0.2in}
We thank Emanuele Castorina and Richard Shaw for useful discussions
and comments on an early version of this paper.
N.P. thanks Laura Newburgh for useful conversations.
We thank the referee for detailed comments on the paper.
M.W. is supported by DOE. N.P. is supported in part by DOE DE-SC0008080. 
This work made extensive use of the NASA Astrophysics Data System and
of the {\tt astro-ph} preprint archive at {\tt arXiv.org}.  The
analysis made use of the computing resources of the National Energy
Research Scientific Computing Center.

\appendix

\section{Signal to noise ratio}
\label{app:noise}

We present a self-contained derivation of the instrument noise power spectrum,
converted to cosmological units. Our derivation is similar to that in
\citet{Alonso2017}, but related expressions have also appeared in
\citet{Whi99,ZalFurHer04,McQ06,Seo2010,Bul15,SeoHir16,Wol17}.

The brightness temperature, $T_b$, is defined in terms of the intensity at
frequency $\nu$ as $I_\nu=2k_BT_b(\nu/c)^2=2k_BT_b/\lambda^2$.
We begin by noting that if we normalize our visibilities in terms of
temperature (rather than intensity) the power can be written in terms of
the brightness temperature power spectrum and window function as
\begin{equation}
  \left\langle\left| V_i\right|^2\right\rangle
  = \int d^2u\ P_T(\vec{u})W(\vec{u})
  \approx P_T(\vec{u}) \int d^2u\ W(\vec{u})
\end{equation}
with the last approximation holding if the window function is compact and
$P_T$ is smooth. 
Conventionally the beam is normalized to unity at peak,
so its area in the $u-v$ plane integrates to unity and thus the window
function integrates to the inverse area
\begin{equation}
  \int d^2u\ W(\vec{u}) \sim \frac{1}{d^2 u} \qquad ,
\end{equation}
which gives
\begin{equation}
  \left\langle\left| V_i\right|^2\right\rangle
  \approx \frac{P_T(\vec{u})}{d^2 u} \qquad .
\label{eqn:pow2vis}
\end{equation}

It may be helpful to derive Eq.~(\ref{eqn:pow2vis}) differently.
If we treat the visibility as measuring a single Fourier mode of the 2D
brightness temperature field, we can relate this to the 2D power spectrum
of this field
\begin{equation}
  \langle V(\vec{\ell}) V^{\star}(\vec{\ell}') \rangle =
  (2\pi)^2 \delta^D(\vec{\ell} - \vec{\ell}')
  P_T(\vec{\ell}=2\pi\vec{u})
  \approx \delta^K_{\vec{\ell},\vec{\ell}'}\frac{P_T(\vec{u})}{d^2 u}
  \qquad ,
\end{equation}
where $\delta^{D,K}$ are the Dirac and Kronecker $\delta$ functions.
The above equation explicitly relates $\vec{u}$ to the 2D
wavevector $\vec{\ell}$, and the last approximation comes from assuming a
discretized set of wavevectors\footnote{For instance, this is exactly what
  happens on an FFT grid in a simulation.}.

In the same units the visibility noise is diagonal \citep{TMS}
\begin{equation}
  \left\langle\left| N_i\right|^2\right\rangle
  = \left[\frac{2k_B}{\lambda^2}\right]^{-2}
    \left[\frac{2k_B T_{\rm sys}}{A_e}\right]^2
    \frac{1}{\Delta\nu\,t_{\rm p}}
  = \left[\frac{\lambda^2\,T_{\rm sys}}{A_e}\right]^2
    \frac{1}{\Delta\nu\,t_{\rm p}}
\end{equation}
per baseline.  Here $T_{\rm sys}$ is the system temperature, $A_e$ the
effective area of the telescope (equal to the aperture efficiency times the
physical area), $\Delta\nu$ is the bandwidth, $t_{\rm p}$ is the
observing time per pointing and we have assumed a single polarization.

The above is all that is needed to implement a matched filter on the data,
where we can work at the level of the visibilities. It is however useful
to translate this into the cosmological units used in the paper. We start
by defining the number of baselines per unit area in the $u-v$ plane,
$n(\vec{u})$, normalized such that
\begin{equation}
  \int d^2u \, n(\vec{u}) = N_{\rm pairs} = \frac{N_a (N_a-1)}{2}
\end{equation}
where $N_a$ is the number of antennae, and $N_{\rm pairs}$ is the number
of pairs (i.e.~instantaneous baselines).
Averaging over the number of baselines, the noise becomes
$\langle N_{i}^2 \rangle/(n(\vec{u}) d^2u)$.
Using Eq.~(\ref{eqn:pow2vis}) we obtain
\begin{equation}
  P_{N}(\vec{u}) = \left[\frac{\lambda^2\,T_{\rm sys}}{A_e}\right]^2 \frac{1}{n(\vec{u})}
    \frac{1}{\Delta\nu\,t_{\rm p}} \, = 
  \left[\frac{\lambda^2\,T_{\rm sys}^2}{A_e}\right] \frac{1}{n(\vec{u})}
    \frac{4\pi f_{\rm sky}}{\Delta\nu\,t_{\rm obs}} \,.
\end{equation}
The last equality follows from $N_{p} \Omega_{p} = 4 \pi f_{\rm sky}$ where
$N_{p}=t_{\rm obs}/t_{\rm p}$ is the number of pointings and $t_{\rm obs}$
is the total observing time.
The area covered by each pointing $\Omega_p$ is approximately given by
$\lambda^2/A_e$. Physically, the above equations assume that each pointing yield
a disjoint set of modes.

To convert this visibility noise into a cosmological power spectrum, we
divide by the mean cosmological brightness temperature \citep{Seo2010}
\begin{equation}
  \bar{T} = 188 \frac{x_{HI}(z) \Omega_{H,0}h(1+z)^2}{H(z)/H_0} \,{\rm mK} \,,
\end{equation}
with $x_{HI}$ the neutral hydrogen fraction, and convert from $\vec{u}$
to $\vec{k}_\perp$ in comoving coordinates and similarly for frequency to
$k_\parallel$ to obtain
\begin{equation}
  P_N = \left( \frac{T_{\rm sys}}{\bar{T}} \right)^2
    \left( \frac{\lambda^2}{A_{\rm e}} \right)
    \frac{4 \pi f_{\rm sky}}{t_{\rm obs} n(\vec{u})} 
    \frac{d^2V}{d\Omega\,d\nu}
\end{equation}
where in a spatially flat model
\begin{equation}
  \frac{d^2V}{d\Omega\,d\nu} = \chi^2\,\frac{d\chi}{dz}\,\frac{dz}{d\nu}
  = \chi^2\,\frac{c\,(1+z)^2}{H(z)\,\nu_0}
\end{equation}
with $\nu_0=1420\,$MHz.
Unfortunately the value of $\Omega_{H,0}\,h$ is quite uncertain and it
enters quadratically in the noise power spectrum.
\citet{Rao06} measure $10^3\Omega_{H,0}\simeq 0.9\pm 0.3$ at $z\approx 1$
through the abundance of damped Lyman-$\alpha$ systems
(see also the compilations of data in \citealt{Pad15,Cri15}).
The measurement of $\Omega_{H,0}b$ through $21\,$cm auto-correlations by
\citet{Swi13} has a similar value and fractional error.
We will consider the range $(0.6-1.2)\times 10^{-3}$ or
$\Omega_{H,0}\,h=(4-9)\times 10^{-4}$.
For $\Omega_{H,0}\,h=4\times 10^{-4}$ and the HIRAX-like interferometer
described in the text operating for 3 years we obtain
$P_{N}\approx 600\,h^{-3}\,{\rm Mpc}^3$ at $z=1$ and $k_\perp=0.2\,h^{-1}$Mpc.
Comparing to the linear matter spectrum, and assuming $b=1$, we have
$(P_L/P_N)(k_{\perp})\approx 1$.
If $\Omega_{H,0}\,h=9\times 10^{-4}$ we obtain
$P_{N}\approx 150\,h^{-3}\,{\rm Mpc}^3$
and have $(P_L/P_N)(k_{\perp})\approx 4$.

While this is similar in spirit to $nP$ in galaxy surveys, it is worth
emphasizing that this quantity is intrisincally 2D, while $nP$ is spherically symmetric.
In particular, at fixed $k$, the average value of $k_\perp$ is $(\pi/4)k$.

\section{Transit Telescopes and the $m$-mode formalism}
\label{app:transit}

The interferometers for 21cm intensity mapping experiments
are designed to be transit telescopes, using the Earth's rotation
to map large areas of the sky. This mapping process simultaneously
performs two operations that are traditionally treated separately -
filling in the $u-v$ plane\footnote{As we discuss later in this section,
  a more appropriate basis for discussing these telescopes are spherical
  harmonics. We use the $u-v$ plane here to mean an appropriate ``Fourier''
  transform of the sky.},
and improving the resolution in the $u-v$ plane\footnote{Recall that a
  single visibility measurement is smeared in the ``$u-v$'' plane by
  the Fourier transform of the primary beam and ``mosaicking'' 
  combines observations of different areas of the sky to make this
  window function more compact.}
by ``mosaicking''. Furthermore, some upcoming experiments,
notably CHIME\footnote{http://chime.phas.ubc.ca/}
and Tianlai\footnote{http://tianlai.bao.ac.cn}, use a close-packed array of
cylinders rather than traditional dishes.
In the CHIME configuration, 4 cylinders (each $20\,$m in diameter and
$\sim 100\,$m long, oriented north-south) are placed adjacent in the east-west
direction \citep{CHIME14}.
The primary beam from such a configuration is highly extended in the
north-south direction, while being focussed by the cylinders in the
east-west direction.

Both of these features naturally cover large angles on the sky. The
natural basis for describing these telescopes is not the usual Fourier
basis, but rather spherical harmonics. However, most astrophysical
signals (including the voids discussed here)
cover small areas in the sky and are easily described
in a flat-sky limit. The goal of this Appendix is to make the connection
between the wide-angle and flat sky formalism explicit. 

We start with a review of the $m$-mode formalism, following \citet{Sha14,Sha15}
who state the fundamental visibility measurements in a spherical harmonic
basis.
We then take the flat-sky limit of this result and show that we recover the
traditional $u-v$ plane interpretation. Making this connection also allows us
to explicitly see how the Earth's rotation fills in the $u-v$ plane. We then develop
the matched filter formalism in this basis. We conclude with a worked example
of the $m$-mode formalism, to help build intuition.

\subsection{Review of the $m$-mode formalism}

Following \citet{Sha14,Sha15}, if the beam transfer function pointed
at azimuth $\phi$ is
\begin{equation}
  B_{ij}(\vechat{n};\phi)\propto A^2(\vechat{n};\phi)
  \,\exp\left[2\pi i\,\vechat{n}\cdot\vec{u}_{ij}(\phi)\right]
\end{equation}
then
\begin{equation}
  V_{ij}(\phi) = \int d\vechat{n}\ T(\vechat{n}) B_{ij}(\vechat{n};\phi)
\end{equation}
(plus noise, of course). We remind the reader to distinguish
between the pointing center of the beam (the azimuth of which
is $\phi$) and the coordinate that integrates over the beam ($\vechat{n}$).
Expanding $T$ and $B_{ij}(\phi)$ into spherical harmonics
\begin{eqnarray}
  T(\vechat{n}) &=& \sum_{\ell m} a_{\ell m}\,Y_{\ell m}(\vechat{n}) \\
  B_{ij}(\vechat{n};\phi) &=& \sum_{\ell m} B_{\ell m}^{ij}
  \,Y_{\ell m}^\star(\vechat{n})
\end{eqnarray}
we obtain
\begin{equation}
  V_{ij}(\phi) = \sum_{\ell m} B_{ij}^{\ell m}(\phi) a_{lm} \,.
\end{equation}
The rotation of the Earth in $\phi$ causes the beam to transform
as $B^{\ell m}(\phi) = B^{\ell m}(0) e^{i m \phi}$. Defining 
\begin{equation}
  V_{ij}^m = \int \frac{d\phi}{2\pi} e^{-im\phi}  V_{ij}(\phi)
\end{equation}
we obtain
\begin{equation}
  V_{ij}^m = \sum_\ell B_{ij}^{\ell m} a_{\ell m}
\end{equation}
where $B_{ij}^{\ell m}$ without an explicit argument
is understood to be at $\phi=0$ (the phase factor cancels out its
conjugate in the definition of $V_{ij}^m$.)
These $V_{ij}^{m}$ (or their Fourier conjugate $V_{ij}(\phi)$
are the fundamental observables of the telescope.

\subsection{The Flat-Sky Approximation}

It is illuminating to show that the above expression recovers the usual
flat-sky Fourier representation for small areas of the sky. We will use
$\vec{\ell}$ to represent the 2D Fourier wavevector, with magnitude $\ell$
and polar angle $\varphi_{\ell}$ (not to be confused with the pointing center
$\phi$). The correspondence between $a_{\ell m}$ and $a(\vec{\ell})$ is
\citep{Whi99,Dat07}
\begin{equation}
  a(\vec{\ell}) = \sqrt{ \frac{4 \pi}{2 \ell +1}} \sum_{m}
  i^{-m} a_{\ell m} e^{i m \varphi_\ell}
\end{equation}
and
\begin{equation}
  a_{\ell m} = \sqrt{ \frac{2 \ell +1}{4 \pi}} i^m
  \int \frac{d\varphi_{\ell}}{2\pi}\ a(\vec{\ell})\, e^{-i m \varphi_\ell} \,.
\end{equation}
with a similar expansion for $B_{ij}^{\ell m}$.
Substituting into the visibility equation,
$V_{ij}(\phi) = \sum_{\ell m} B_{ij}^{\ell m}(\phi) a_{\ell m}$,
we obtain, for large $\ell$,
\begin{equation}
  V_{ij}(\phi) \approx \frac{1}{(2\pi)^3} \sum_{\ell m}
  \int d\varphi_{\ell} d\varphi_{\ell'}\, \ell a(\vec{\ell})
  B(\vec{\ell}', \phi) e^{i m (\varphi_{\ell} - \varphi_{\ell'})} \,,
\end{equation}
where $\vec{\ell}$ and $\vec{\ell}'$ have the same magnitude.
Doing the sum over $m$ yields a $\delta$-function that collapses
one of the azimuthal integrals to yield
\begin{equation}
  V_{ij}(\phi) \approx \int \frac{\ell\,d\ell\,d\varphi_{\ell}}{(2\pi)^2}
  \ a(\vec{\ell}) B(\vec{\ell}, \phi)
\end{equation}
where we have approximated the sum over $\ell$ by an integral.
The above shows that the visibilities approximately measure a mode
$\vec{\ell}$, smeared by the Fourier transform of the beam function. 

We can use the above results to understand how the rotation of the Earth
fills in the $u-v$ plane. In the flat-sky limit, the Fourier transform
of the beam is $B(\vec{\ell})\sim\sum_m i^{-m}B_{\ell m}\exp[im\varphi_{\ell}]$.
Rotating about the $z$ axis by $\alpha$ scales the $B_{\ell m}$ by
$\exp[im\alpha]$, which is clearly equivalent to rotating $\vec{\ell}$ by
$\alpha$. The $u-v$ coverage of the telescope traces out circles in the
$u-v$ plane as the Earth rotates.
We note that this is different from the usual result for interferometers,
and reflects the transit nature of these telescopes.

\subsection{Matched Filters}

In order to define the matched filter, we need to express the signal in
terms of the observable quantities, in this case the visibilities. 
Since all of the objects of interest in this study are
$\mathcal{O}(10\,{\rm Mpc})$ in size, at a distance of $>1\,$Gpc, they
subtend small angles on the sky, allowing us to express the signal
using the same flat-sky Fourier representation used in the main paper.

To begin, consider a single frequency, corresponding to a fixed redshift or
(redshift-space) distance.  Suppose our template, $\tau$, is centered at
$\theta=0$, is $\phi$-independent and non-zero only when $\theta\ll 1$.
We expand
\begin{eqnarray}
  \tau_{\ell m}(\vechat{z}) &=& \int d\vechat{n}\ Y_{\ell m}^\star(\vechat{n})\,\tau(\theta) \\
  &=& 2\pi\,\delta^K_{m0}\,\sqrt{\frac{2\ell+1}{4\pi}}
      \int d(\cos\theta) P_\ell(\cos\theta)\,\tau(\theta) \\
  &\simeq& \delta^K_{m0}\,\sqrt{\frac{2\ell+1}{4\pi}}
      \ \left[2\pi\int \tilde{\omega}\,d\tilde{\omega}
      \ J_0(\ell\tilde{\omega})\,\tau(\tilde{\omega})\right]
\end{eqnarray}
where in the last line we have defined
$\tilde{\omega}=2\sin(\theta/2)\simeq\theta$
and used $P_\ell(\cos\theta)\approx J_0(\ell\theta)$ for $\theta\ll 1$.
The $\sqrt{(2\ell+1)/4\pi}$ is just $Y_{\ell 0}(\vechat{z})$.
If we extend the upper limit of $\tilde{\omega}$-integration to infinity,
we recognize in the brackets on the last line the Hankel transform of $\tau$
or the 2D Fourier transform of $\tau$ with spherical symmetry
(e.g.~\citealt{BonEfs87}).

Now we can rotate the template from the north pole ($\vechat{z}$) to an arbitrary $\vechat{n}$
using Wigner functions, $\mathcal{D}^\ell_{m'm}$.
However, in our case $\tau_{\ell m}\propto\delta^K_{m0}$ and
$Y_{\ell m}\propto\mathcal{D}^\ell_{0m}$ so that the spherical harmonic
coefficients for a template centered on $\vechat{n}$ are
\begin{eqnarray}
  \tau_{\ell m}(\vechat{n}) &=& \sqrt{\frac{4\pi}{2\ell+1}}
  \ Y_{\ell m}^\star(\vechat{n})\ \tau_{\ell 0}(\vechat{z}) \\
  &=& \left[2\pi\int\tilde{\omega}\,d\tilde{\omega}\ J_0(\ell\tilde{\omega})
  \,\tau(\tilde{\omega})\right]\,Y_{\ell m}^\star(\vechat{n})
\end{eqnarray}
(with no implied sum over $\ell$).

These $\tau_{\ell m}$ can now be inserted into our formula for the $m$-mode
visibility to obtain
\begin{equation}
  {\cal V}_{ij}^m (\vechat{n}) = \sum_{\ell} B_{ij}^{\ell m}\,Y_{\ell m}^\star(\vechat{n})
  \left[2\pi\int\tilde{\omega}\,d\tilde{\omega}\ J_0(\ell\tilde{\omega})
  \,\tau(\tilde{\omega})\right]
\end{equation}
This is the central relation needed for the matched filter, as it expresses
a linear relationship between the observable and the template.
We recognize the the combination $B^{\ell m}_{ij}\,Y_{\ell m}^\star$ as the
beam transfer function, $B_{ij}(\vechat{n};\phi)$, evaluated at the position of
the object but now modulated by the Fourier transform of $\tau$.

The above expressions are all for a single frequency.
If we now perform the Fourier transform in frequency, the term in square
brackets becomes the 3D Fourier transform for an azimuthally symmeteric
function in cylindrical coordinates: $\tau(k_\perp,k_\parallel)$ with
$\ell\simeq |k_\perp|$.
For a narrow range of frequencies (corresponding to an astrophysical object
such as a void or protocluster for example) the $k_\perp$ probed by the
interferometer are almost constant.  For a wide range of frequencies one
must account for the shifting of $\vec{u}_{ij}$, and $\ell$, with wavelength
at fixed baseline separation.
This represents no difficulty in principle, since we need only evaluate
our template where there is data, but it formally breaks the Fourier transform
property. It is important to note that this Fourier transform is not necessary for
the matched filter, which can be written in visibility-frequency space.

As before, the matched filter is defined by
\begin{equation}
  \hat{A}(\vechat{n}) = \frac{V_{ij}^m N^{-1} {\cal V}_{ij}^m (\vechat{n})}{{\cal V}_{ij}^m(\vechat{n}) N^{-1} {\cal V}_{ij}^m(\vechat{n})} \,,
\end{equation}
where the noise covariance matrix both includes the visibility noise and projects out contaminated
modes. There are a few practical differences between this treatment and the flat-sky Fourier version
we discuss in the main text. In the simplified flat sky treatment, shifting the matched filter
to an arbitrary position $\vec{x}$ was simply a multiplication of $\hat{A}$ by $\exp(i \vec{k} \vec{r})$, which allowed
us to efficiently evaluate the matched filter at all possible void positions with inverse FFTs.
In particular, the denominator of $\hat{A}$ is translation-invariant. While these simplifications remain
true in the azimuthal direction, they no longer hold for the polar or radial directions. Therefore, one
must explicitly evaluate the matched filter at all possible void positions. It may be possible to reduce
the computational burden by using the Fourier versions of the expressions about more sparsely sampled
central void positions. Since the precise implementation will be survey dependent, we do not pursue
more detailed implementations here.

\subsection{A Worked Example}

We conclude with an analytic example to make this formalism more
concrete. Our discussion here parallels that in \citet{BunWhi07}.
Consider the interferometer situated at the equator ($\theta_0=\pi/2$, $\phi_0=0$)
and looking directly overhead.
The baselines, $\vec{u}_{ij}$, lie in the $y-z$ plane.  We will consider two
cases, a north-south baseline ($\vec{u}=u\vechat{z}$) and an east-west baseline
($\vec{u}=u\vechat{y}$).  For a small field of view, we approximate the
sky as flat with Cartesian coordinates $\phi, \delta$, where $\delta \equiv \pi/2-\theta$
is the latitude.
A Gaussian beam, normalized to unit peak, then has
\begin{equation}
  B(\vechat{n})=B(\phi,\delta) =
  \exp\left[-\frac{\phi^2+\delta^2}{2\sigma^2}\right]\times
  \left\{\begin{array}{cc}
  \exp[2\pi i u \phi]   & {\rm for}\ \vechat{y} (\,{\rm EW})\\
  \exp[2\pi i u \delta] & {\rm for}\ \vechat{z} (\,{\rm NS})\end{array}
  \right.
\end{equation}
where we have suppressed the $ij$ indices labeling the visibility for
convenience.

The visibility for this baseline is
\begin{equation}
  V(\vechat{n}) = \int d\vechat{n}\ B(\vechat{n}) T(\vechat{n}) \,.
\end{equation}
Instead of immediately going to the spherical harmonic expansion, it is
algebraicly illuminating and amusing to imagine the sky as a torus. The
appropriate orthogonal basis is then the usual Fourier basis
\begin{equation}
  V = \int d\vechat{n} \sum B_{nm} e^{-i n \delta} e^{-i m \phi}
  \sum T_{n'm'} e^{i n' \delta} e^{i m' \phi}
\end{equation}
which collapses to
\begin{equation}
  V^{m} = (2\pi)^2 \sum_{nm} B_{nm} T_{nm}
\end{equation}
where we have also implicitly gone to the $m$-mode basis (to account for the
Earth's rotation). This expression is analogous to the spherical harmonic version.
The beam multipole moments
are then given by
\begin{equation}
  B_{nm} = \int \frac{d\phi\,d\delta}{(2\pi)^2}
  \ B(\phi, \delta)\ e^{i n \delta} e^{i m \phi} \,.
\end{equation}
Since we assume the beams are compact in both $\phi$ and $\delta$,
we are free to extend the limits of integration to $\pm \infty$. 
For the specific case of our Gaussian beam, these integrals are then
just Gaussian integrals and can be easily evaluated.
For an EW baseline, we get
\begin{equation}
  B_{nm} \propto \exp\left[-\frac{\sigma^2 n^2}{2}\right]
                 \exp\left[-\frac{\sigma^2 (m \pm 2\pi u)^2}{2}\right]
\end{equation}
while for the NS baseline, we find
\begin{equation}
  B_{nm} \propto \exp\left[-\frac{\sigma^2 (n \pm 2\pi u)^2}{2}\right]
                 \exp\left[-\frac{\sigma^2 m^2}{2}\right]
\end{equation}
where the $\pm$ cases come from the two possible choices for the sign of $u$.
These have a clear physical interpretation - the EW baseline probes modes
centered around $(n=0, m=\pm 2\pi u)$ while the NS baseline is centered on
$(n=2\pi u, m=0)$.  Note that these expressions indicate that it is the
baseline distribution and the primary beam which delineate the range of
$(\ell m)$ modes which need to be kept in the sums of the previous section.

Returning to a spherical sky, we will adopt a similar strategy to understand
what modes a given baseline probes.  Since the beam is compact, we will
approximate the spherical harmonics by a Fourier series, in which case the
algebra proceeds as in the case of the torus.
All that will remain will be to understand the correspondence between mode
coefficients $n$ on the torus and $(\ell, m)$ on the sphere\footnote{Note
that in the $\phi$ direction, both the sphere and the torus have Fourier
expansions.}.

For our specific case, the multipole moments then become
\begin{eqnarray}
  B_{\ell m} &=& \int d\phi\,d\sin\delta
  \ Y_{\ell m}\left(\frac{\pi}{2}-\delta, \phi\right)B(\phi,\delta) \\
  &\simeq& \int_{-\infty}^{\infty} d\phi\,d\delta
  \ Y_{\ell m}\left(\frac{\pi}{2}-\delta, \phi\right)B(\phi,\delta)
\end{eqnarray}
where we assume $\delta \ll 1$ in the second line. Near the equator, we have
\footnote{The approximation agrees to the first two terms in the Taylor series.
  For completeness, we note that
  \begin{equation}
    N_{\ell m} = 2^{m} \sqrt{\pi} \sqrt{\frac{2\ell +1}{4\pi}}
    \sqrt{\frac{(\ell - m)!}{(\ell + m)!}}  \,
    \frac{1}{\Gamma \left(\frac{1}{2}-\frac{(l+m)}{2}\right)
      \Gamma \left(1+ \frac{(l-m)}{2}\right)} .
  \end{equation}
}
\begin{equation}
  Y_{\ell m} \simeq N_{\ell m} e^{i m \phi}
  \left\{\begin{array}{cc}
  \hphantom{-}\cos n_{\ell m} \delta & {\rm for}\ \ell+m\ {\rm even} \\
  -\sin n_{\ell m} \delta & {\rm for}\ \ell+m\ {\rm odd} \end{array}
  \right.,
\end{equation}
where $N_{\ell m}$ is a constant and
\begin{equation}
  n_{\ell m}^2 = \ell (\ell+1) -m^2 -
  \left\{\begin{array}{cc}
  0 & {\rm for}\ \ell+m\ {\rm even} \\
  1 & {\rm for}\ \ell+m\ {\rm odd} \end{array}
  \right..
\end{equation}

Since we have reduced the problem to the toroidal sky case, we proceed as
before and find that EW baselines measure modes centered on
$(n_{\ell m}=0, m=2 \pi u)$.
In the limit that $\ell \gg 1$, this implies that these baselines measure
modes with $m \sim 2 \pi u$, $\ell \sim m$.
As one might expect, $\ell$ and $m$ are coupled together by the spherical
geometry.
For NS baselines, the $m$-mode visibilities probe $(n_{\ell m}=2 \pi u, m=0)$
or $\ell \sim 2\pi u$, $m \sim 0$.
The azimuthal symmetry of the baseline configuration is reflected in the
visibilities isolating the $m \sim 0$ modes.
These two cases represent the two limiting cases; baselines with components
in both the EW and NS directions will probe more general $\ell, m$ modes.

For this particular case, this also completes the correspondence with the usual
flat-sky treatment where a baseline measures a particular $\vec{\ell}$ Fourier
mode.  Here, the visibility $m$ modes measure particular $\ell, m$ modes.

\section{Modeling the $21\,$cm signal}
\label{app:HI}

In the main text we have assumed that neutral hydrogen traces the mass
field in an unbiased manner for the purposes of testing our matched filter
on simulations.  In this appendix we present a more refined model and argue
that this assumption is conservative (for our purposes).

At low $z$ most of the hydrogen in the Universe is ionized, and the $21\,$cm
signal comes only from self-shielded regions such as galaxies\footnote{Most
likely between the outskirts of disks until where the gas becomes molecular
within star-forming regions.}.  Unfortunately there are not many observational
constraints on the manner in which HI traces galaxies and halos in the high-$z$
Universe.  There have been a large number of approaches to modeling this
uncertain signal.
Some approaches work directly at the level of the density field.
For example, \citet{Sha14,Sha15} use Gaussian density fields.
\citet{Bul15} assumes a constant bias times the matter power spectrum
(this is implicitly what we do in the main text, with $b=1$).
The CRIME code by \citet{AloFerSan14} uses lognormal realizations.
\citet{BagWhi03} selected dark matter particles based on a density threshold
to mock up self-shielded regions.

An alternative is to use a halo-based approach, specifying the mass of
HI to assign to a dark matter halo of a given mass, $M_h$.  A popular
model was introduced by \citet{Bag10}, which populated halos with circular
velocities above $30\,$km/s with HI such that the HI mass saturates at high
halo mass.  A similar model was proposed by \citet{BarHae10,BarHae14}, who
modeled the low-$M$ cutoff as an exponential.
\citet{Mar10} use abundance matching between blue galaxies in the HI mass
function at $z\approx 0$.
\citet{Gon11} employ a double power-law model.
\citet{See16} propose a form with an exponential cut-off at both low and
high halo masses.
\citet{PadRef17} allow a non-unity slope in addition to the high and low
mass cut-offs.
The model we shall follow is due to \citet{CasVil16}, which assumes
\begin{equation}
  M_{HI} \propto M_h^\alpha\ e^{-M_{\rm cut}/M_h}
\label{eqn:HI_HOD}
\end{equation}
with the constant of proportionality adjusted to match the observed value
of $\Omega_{HI}$.  Aside from the normalization, this model has two free
parameters, $\alpha$ and $M_{\rm cut}$, which control the behavior at high
and low halo masses.  There is evidence from simulations that $\alpha<1$
\citep[e.g.][]{Dav13,Vil16} with $\alpha\approx 3/4$ a reasonable estimate.
We shall use this value.  Note that in contrast to some of the other models
this assumption puts significant HI mass in higher mass halos.
There is some evidence at $z\simeq 0$ that HI is depleted in galaxies within
clusters \citep[e.g.][]{Sol01}, but the behavior at $z\sim 1$ is unknown.
In the simulations of \citet{CasVil16} the trend of $M_{HI}$ with $M_h$
is different at high and low redshift.
The remaining free parameter, $M_{\rm cut}$, then adjusts the bias\footnote{For
$\alpha=3/4$ at $z\approx 1$ the bias ranges from $1.4$ to $1.7$ as
lg$M_{\rm cut}$ runs from $10.5$ to $11.5$ in $h^{-1}M_\odot$ units.
This is consistent with the amplitude of the measured clustering at $z\sim 1$
by \citet{Cha10,Swi13} but those measurements are not precise enough to
place strong limits on the bias.} of the HI.
While a range of values are allowed within the observational constraints,
typical values for the low-mass cut-off, $M_{\rm cut}$, are around
$10^{11}\,h^{-1}M_\odot$.  We shall explore a range around this value
(lg$M_{\rm cut}=10.5$, 11 and $11.5$ with masses in $h^{-1}M_\odot$)
to illustrate the effects.

The simulations used in the main body of this paper do not have sufficient
resolution to track the halos expected to host much of the HI at $z\sim 1$.
Thus in this appendix we use a different simulation, run with the same code,
which employed $2560^3$ particles in a box of side $256\,h^{-1}$Mpc.
This is the same simulation as used in \citet{Sta15a,Sta15b}, to which the
reader is referred for more details.
We generate a mock HI field from the $z\simeq 1$ halo catalog using the
mapping of Eq.~(\ref{eqn:HI_HOD}).

We find voids in this simulation using the same technique as described in
the main text.  For completeness we also find protoclusters, in a manner
similar to \citet{Sta15a}: starting from a friends-of-friends halo catalog
(with a linking length of $0.168$ times the mean interparticle spacing) we
select each $z=0$ halo more massive than $10^{14}\,h^{-1}M_\odot$.
We then track the particles within a few hundred kpc of the most bound
particle back to $z=1$.  The center of mass of these is taken to be the
protocluster position at $z=1$.

\begin{figure}
\begin{center}
\resizebox{\columnwidth}{!}{\includegraphics{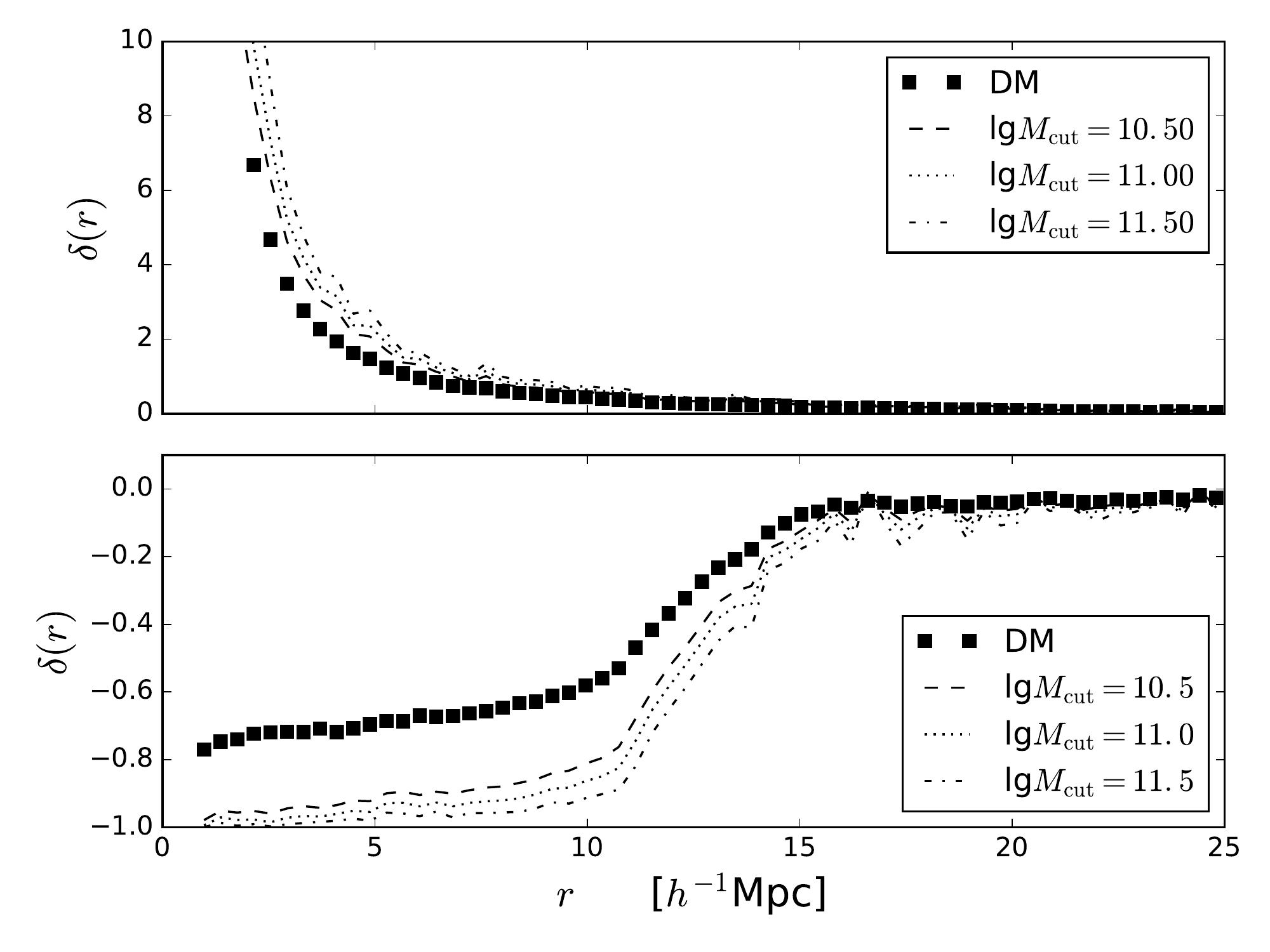}}
\caption{The (real-space) profiles of voids and protoclusters at $z\simeq 1$,
as in Fig.~\ref{fig:profile1d}.
The upper panel shows the $M>10^{14}\,h^{-1}M_\odot$ protocluster profile
in the mass and in the HI for three values of the cut-off mass in
Eq.~(\protect\ref{eqn:HI_HOD}), specified as $\log_{10}$ of the mass in
$h^{-1}M_\odot$.  The lower panel shows the same comparison for voids of
$10<r_s<15\,h^{-1}$Mpc.}
\label{fig:prof_HI}
\end{center}
\end{figure}

A comparison of the (real-space) profiles of protoclusters and voids in
the dark matter and mock HI at $z\simeq 1$ is shown in Fig.~\ref{fig:prof_HI}
for three values of $M_{\rm cut}$.
The curves are noisier than from the larger volume simulations, due to the
poorer statistics, however we see that the protoclusters in the HI have just
as much broad, distributed emission as the matter profiles.
The voids in the HI have a qualitatively similar ``bucket shaped'' profile
to the mass density, but are notably more empty.
As noted by \citet{TinCon09}, the halo mass function shifts dramatically
to lower masses in underdense regions.  Thus we expect to see voids in the
massive halo and HI distributions be ``more empty'' than in the mass.
Given the greater contrast in HI than in the matter, our approximation in the
main text is conservative from the point of view of finding protoclusters and
voids with $21\,$cm experiments.

\bibliographystyle{mn2e}
\bibliography{ms}

\begin{thebibliography}{}
\makeatletter
\relax
\def\mn@urlcharsother{\let\do\@makeother \do\$\do\&\do\#\do\^\do\_\do\%\do\~}
\def\mn@doi{\begingroup\mn@urlcharsother \@ifnextchar [ {\mn@doi@}
  {\mn@doi@[]}}
\def\mn@doi@[#1]#2{\def\@tempa{#1}\ifx\@tempa\@empty \href
  {http://dx.doi.org/#2} {doi:#2}\else \href {http://dx.doi.org/#2} {#1}\fi
  \endgroup}
\def\mn@eprint#1#2{\mn@eprint@#1:#2::\@nil}
\def\mn@eprint@arXiv#1{\href {http://arxiv.org/abs/#1} {{\tt arXiv:#1}}}
\def\mn@eprint@dblp#1{\href {http://dblp.uni-trier.de/rec/bibtex/#1.xml}
  {dblp:#1}}
\def\mn@eprint@#1:#2:#3:#4\@nil{\def\@tempa {#1}\def\@tempb {#2}\def\@tempc
  {#3}\ifx \@tempc \@empty \let \@tempc \@tempb \let \@tempb \@tempa \fi \ifx
  \@tempb \@empty \def\@tempb {arXiv}\fi \@ifundefined
  {mn@eprint@\@tempb}{\@tempb:\@tempc}{\expandafter \expandafter \csname
  mn@eprint@\@tempb\endcsname \expandafter{\@tempc}}}

\bibitem[\protect\citeauthoryear{{Ali} \& {Bharadwaj}}{{Ali} \&
  {Bharadwaj}}{2014}]{ORT14}
{Ali} S.~S.,  {Bharadwaj} S.,  2014, \mn@doi [Journal of Astrophysics and
  Astronomy] {10.1007/s12036-014-9301-1}, \href
  {http://adsabs.harvard.edu/abs/2014JApA...35..157A} {35, 157}

\bibitem[\protect\citeauthoryear{{Alonso}, {Ferreira}  \& {Santos}}{{Alonso}
  et~al.}{2014}]{AloFerSan14}
{Alonso} D.,  {Ferreira} P.~G.,   {Santos} M.~G.,  2014, \mn@doi [\mnras]
  {10.1093/mnras/stu1666}, \href
  {http://adsabs.harvard.edu/abs/2014MNRAS.444.3183A} {444, 3183}

\bibitem[\protect\citeauthoryear{{Alonso}, {Ferreira}, {Jarvis}  \&
  {Moodley}}{{Alonso} et~al.}{2017}]{Alonso2017}
{Alonso} D.,  {Ferreira} P.~G.,  {Jarvis} M.~J.,   {Moodley} K.,  2017,
  preprint, \href {http://adsabs.harvard.edu/abs/2017arXiv170401941A} {}
  (\mn@eprint {arXiv} {1704.01941})

\bibitem[\protect\citeauthoryear{{Ansari} et~al.,}{{Ansari}
  et~al.}{2012}]{BAOradio12}
{Ansari} R.,  et~al., 2012, \mn@doi [\aap] {10.1051/0004-6361/201117837}, \href
  {http://adsabs.harvard.edu/abs/2012A%26A...540A.129A} {540, A129}

\bibitem[\protect\citeauthoryear{{Bagla} \& {White}}{{Bagla} \&
  {White}}{2003}]{BagWhi03}
{Bagla} J.~S.,  {White} M.,  2003, in {Ikeuchi} S.,  {Hearnshaw} J.,   {Hanawa}
  T.,  eds,  Astronomical Society of the Pacific Conference Series Vol. 289,
  The Proceedings of the IAU 8th Asian-Pacific Regional Meeting, Volume 1. pp
  251--254 (\mn@eprint {} {astro-ph/0212228})

\bibitem[\protect\citeauthoryear{{Bagla}, {Khandai}  \& {Datta}}{{Bagla}
  et~al.}{2010}]{Bag10}
{Bagla} J.~S.,  {Khandai} N.,   {Datta} K.~K.,  2010, \mn@doi [\mnras]
  {10.1111/j.1365-2966.2010.16933.x}, \href
  {http://adsabs.harvard.edu/abs/2010MNRAS.407..567B} {407, 567}

\bibitem[\protect\citeauthoryear{{Banerjee} \& {Dalal}}{{Banerjee} \&
  {Dalal}}{2016}]{BanDal16}
{Banerjee} A.,  {Dalal} N.,  2016, \mn@doi [\jcap]
  {10.1088/1475-7516/2016/11/015}, \href
  {http://adsabs.harvard.edu/abs/2016JCAP...11..015B} {11, 015}

\bibitem[\protect\citeauthoryear{{Barnes} \& {Haehnelt}}{{Barnes} \&
  {Haehnelt}}{2010}]{BarHae10}
{Barnes} L.~A.,  {Haehnelt} M.~G.,  2010, \mn@doi [\mnras]
  {10.1111/j.1365-2966.2009.16172.x}, \href
  {http://adsabs.harvard.edu/abs/2010MNRAS.403..870B} {403, 870}

\bibitem[\protect\citeauthoryear{{Barnes} \& {Haehnelt}}{{Barnes} \&
  {Haehnelt}}{2014}]{BarHae14}
{Barnes} L.~A.,  {Haehnelt} M.~G.,  2014, \mn@doi [\mnras]
  {10.1093/mnras/stu445}, \href
  {http://adsabs.harvard.edu/abs/2014MNRAS.440.2313B} {440, 2313}

\bibitem[\protect\citeauthoryear{{Battye}, {Davies}  \& {Weller}}{{Battye}
  et~al.}{2004}]{Bat04}
{Battye} R.~A.,  {Davies} R.~D.,   {Weller} J.,  2004, \mn@doi [\mnras]
  {10.1111/j.1365-2966.2004.08416.x}, \href
  {http://adsabs.harvard.edu/abs/2004MNRAS.355.1339B} {355, 1339}

\bibitem[\protect\citeauthoryear{{Bond} \& {Efstathiou}}{{Bond} \&
  {Efstathiou}}{1987}]{BonEfs87}
{Bond} J.~R.,  {Efstathiou} G.,  1987, \mn@doi [\mnras]
  {10.1093/mnras/226.3.655}, \href
  {http://adsabs.harvard.edu/abs/1987MNRAS.226..655B} {226, 655}

\bibitem[\protect\citeauthoryear{{Bull}, {Ferreira}, {Patel}  \&
  {Santos}}{{Bull} et~al.}{2015}]{Bul15}
{Bull} P.,  {Ferreira} P.~G.,  {Patel} P.,   {Santos} M.~G.,  2015, \mn@doi
  [\apj] {10.1088/0004-637X/803/1/21}, \href
  {http://adsabs.harvard.edu/abs/2015ApJ...803...21B} {803, 21}

\bibitem[\protect\citeauthoryear{{Bunn} \& {White}}{{Bunn} \&
  {White}}{2007}]{BunWhi07}
{Bunn} E.~F.,  {White} M.,  2007, \mn@doi [\apj] {10.1086/509867}, \href
  {http://adsabs.harvard.edu/abs/2007ApJ...655...21B} {655, 21}

\bibitem[\protect\citeauthoryear{{Cai}, {Padilla}  \& {Li}}{{Cai}
  et~al.}{2015}]{Cai15}
{Cai} Y.-C.,  {Padilla} N.,   {Li} B.,  2015, \mn@doi [\mnras]
  {10.1093/mnras/stv777}, \href
  {http://adsabs.harvard.edu/abs/2015MNRAS.451.1036C} {451, 1036}

\bibitem[\protect\citeauthoryear{{Cai}, {Taylor}, {Peacock}  \&
  {Padilla}}{{Cai} et~al.}{2016}]{Cai16}
{Cai} Y.-C.,  {Taylor} A.,  {Peacock} J.~A.,   {Padilla} N.,  2016, \mn@doi
  [\mnras] {10.1093/mnras/stw1809}, \href
  {http://adsabs.harvard.edu/abs/2016MNRAS.462.2465C} {462, 2465}

\bibitem[\protect\citeauthoryear{{Castorina} \&
  {Villaescusa-Navarro}}{{Castorina} \& {Villaescusa-Navarro}}{2016}]{CasVil16}
{Castorina} E.,  {Villaescusa-Navarro} F.,  2016, preprint, \href
  {http://adsabs.harvard.edu/abs/2016arXiv160905157C} {} (\mn@eprint {arXiv}
  {1609.05157})

\bibitem[\protect\citeauthoryear{{Chang}, {Pen}, {Bandura}  \&
  {Peterson}}{{Chang} et~al.}{2010}]{Cha10}
{Chang} T.-C.,  {Pen} U.-L.,  {Bandura} K.,   {Peterson} J.~B.,  2010, \mn@doi
  [\nat] {10.1038/nature09187}, \href
  {http://adsabs.harvard.edu/abs/2010Natur.466..463C} {466, 463}

\bibitem[\protect\citeauthoryear{{Chen}}{{Chen}}{2012}]{Tianlai12}
{Chen} X.,  2012, in International Journal of Modern Physics Conference Series.
  pp 256--263 (\mn@eprint {arXiv} {1212.6278}),
  \mn@doi{10.1142/S2010194512006459}

\bibitem[\protect\citeauthoryear{{Chiang}, {Overzier}, {Gebhardt}  \&
  {Henriques}}{{Chiang} et~al.}{2017}]{Chi17}
{Chiang} Y.-K.,  {Overzier} R.~A.,  {Gebhardt} K.,   {Henriques} B.,  2017,
  preprint, \href {http://adsabs.harvard.edu/abs/2017arXiv170501634C} {}
  (\mn@eprint {arXiv} {1705.01634})

\bibitem[\protect\citeauthoryear{{Clampitt}, {Cai}  \& {Li}}{{Clampitt}
  et~al.}{2013}]{Cla13}
{Clampitt} J.,  {Cai} Y.-C.,   {Li} B.,  2013, \mn@doi [\mnras]
  {10.1093/mnras/stt219}, \href
  {http://adsabs.harvard.edu/abs/2013MNRAS.431..749C} {431, 749}

\bibitem[\protect\citeauthoryear{{Cohn}, {White}, {Chang}, {Holder},
  {Padmanabhan}  \& {Dor{\'e}}}{{Cohn} et~al.}{2016}]{Coh16}
{Cohn} J.~D.,  {White} M.,  {Chang} T.-C.,  {Holder} G.,  {Padmanabhan} N.,
  {Dor{\'e}} O.,  2016, \mn@doi [\mnras] {10.1093/mnras/stw108}, \href
  {http://adsabs.harvard.edu/abs/2016MNRAS.457.2068C} {457, 2068}

\bibitem[\protect\citeauthoryear{{Colberg} et~al.,}{{Colberg}
  et~al.}{2008}]{Col08}
{Colberg} J.~M.,  et~al., 2008, \mn@doi [\mnras]
  {10.1111/j.1365-2966.2008.13307.x}, \href
  {http://adsabs.harvard.edu/abs/2008MNRAS.387..933C} {387, 933}

\bibitem[\protect\citeauthoryear{{Crighton} et~al.,}{{Crighton}
  et~al.}{2015}]{Cri15}
{Crighton} N.~H.~M.,  et~al., 2015, \mn@doi [\mnras] {10.1093/mnras/stv1182},
  \href {http://adsabs.harvard.edu/abs/2015MNRAS.452..217C} {452, 217}

\bibitem[\protect\citeauthoryear{{Datta}, {Choudhury}  \& {Bharadwaj}}{{Datta}
  et~al.}{2007}]{Dat07}
{Datta} K.~K.,  {Choudhury} T.~R.,   {Bharadwaj} S.,  2007, \mn@doi [\mnras]
  {10.1111/j.1365-2966.2007.11747.x}, \href
  {http://adsabs.harvard.edu/abs/2007MNRAS.378..119D} {378, 119}

\bibitem[\protect\citeauthoryear{{Dav{\'e}}, {Katz}, {Oppenheimer}, {Kollmeier}
   \& {Weinberg}}{{Dav{\'e}} et~al.}{2013}]{Dav13}
{Dav{\'e}} R.,  {Katz} N.,  {Oppenheimer} B.~D.,  {Kollmeier} J.~A.,
  {Weinberg} D.~H.,  2013, \mn@doi [\mnras] {10.1093/mnras/stt1274}, \href
  {http://adsabs.harvard.edu/abs/2013MNRAS.434.2645D} {434, 2645}

\bibitem[\protect\citeauthoryear{{Dong}, {Pierpaoli}, {Gunn}  \&
  {Wechsler}}{{Dong} et~al.}{2008}]{Don08}
{Dong} F.,  {Pierpaoli} E.,  {Gunn} J.~E.,   {Wechsler} R.~H.,  2008, \mn@doi
  [\apj] {10.1086/522490}, \href
  {http://adsabs.harvard.edu/abs/2008ApJ...676..868D} {676, 868}

\bibitem[\protect\citeauthoryear{{Falck}, {Koyama}, {Zhao}  \&
  {Cautun}}{{Falck} et~al.}{2017}]{Fal17}
{Falck} B.,  {Koyama} K.,  {Zhao} G.,   {Cautun} M.,  2017, preprint, \href
  {http://adsabs.harvard.edu/abs/2017arXiv170408942F} {} (\mn@eprint {arXiv}
  {1704.08942})

\bibitem[\protect\citeauthoryear{{Furlanetto}, {Oh}  \& {Briggs}}{{Furlanetto}
  et~al.}{2006}]{Fur06}
{Furlanetto} S.~R.,  {Oh} S.~P.,   {Briggs} F.~H.,  2006, \mn@doi [\physrep]
  {10.1016/j.physrep.2006.08.002}, \href
  {http://adsabs.harvard.edu/abs/2006PhR...433..181F} {433, 181}

\bibitem[\protect\citeauthoryear{{Gong}, {Chen}, {Silva}, {Cooray}  \&
  {Santos}}{{Gong} et~al.}{2011}]{Gon11}
{Gong} Y.,  {Chen} X.,  {Silva} M.,  {Cooray} A.,   {Santos} M.~G.,  2011,
  \mn@doi [\apjl] {10.1088/2041-8205/740/1/L20}, \href
  {http://adsabs.harvard.edu/abs/2011ApJ...740L..20G} {740, L20}

\bibitem[\protect\citeauthoryear{{Hamaus}, {Sutter}  \& {Wandelt}}{{Hamaus}
  et~al.}{2014}]{Ham14}
{Hamaus} N.,  {Sutter} P.~M.,   {Wandelt} B.~D.,  2014, \mn@doi [Physical
  Review Letters] {10.1103/PhysRevLett.112.251302}, \href
  {http://adsabs.harvard.edu/abs/2014PhRvL.112y1302H} {112, 251302}

\bibitem[\protect\citeauthoryear{{Hamaus}, {Sutter}, {Lavaux}  \&
  {Wandelt}}{{Hamaus} et~al.}{2015}]{Ham15}
{Hamaus} N.,  {Sutter} P.~M.,  {Lavaux} G.,   {Wandelt} B.~D.,  2015, \mn@doi
  [JCAP] {10.1088/1475-7516/2015/11/036}, \href
  {http://adsabs.harvard.edu/abs/2015JCAP...11..036H} {11, 036}

\bibitem[\protect\citeauthoryear{{Hamaus}, {Cousinou}, {Pisani}, {Aubert},
  {Escoffier}  \& {Weller}}{{Hamaus} et~al.}{2017}]{Ham17}
{Hamaus} N.,  {Cousinou} M.-C.,  {Pisani} A.,  {Aubert} M.,  {Escoffier} S.,
  {Weller} J.,  2017, preprint, \href
  {http://adsabs.harvard.edu/abs/2017arXiv170505328H} {} (\mn@eprint {arXiv}
  {1705.05328})

\bibitem[\protect\citeauthoryear{{Hawken} et~al.,}{{Hawken}
  et~al.}{2016}]{Haw17}
{Hawken} A.~J.,  et~al., 2016, preprint, \href
  {http://adsabs.harvard.edu/abs/2016arXiv161107046H} {} (\mn@eprint {arXiv}
  {1611.07046})

\bibitem[\protect\citeauthoryear{{Heitmann} et~al.,}{{Heitmann}
  et~al.}{2008}]{Hei08}
{Heitmann} K.,  et~al., 2008, \mn@doi [Computational Science and Discovery]
  {10.1088/1749-4699/1/1/015003}, \href
  {http://adsabs.harvard.edu/abs/2008CS%26D....1a5003H} {1, 015003}

\bibitem[\protect\citeauthoryear{{Kaiser}}{{Kaiser}}{1987}]{Kai87}
{Kaiser} N.,  1987, \mn@doi [\mnras] {10.1093/mnras/227.1.1}, \href
  {http://adsabs.harvard.edu/abs/1987MNRAS.227....1K} {227, 1}

\bibitem[\protect\citeauthoryear{{Kochanek}, {White}, {Huchra}, {Macri},
  {Jarrett}, {Schneider}  \& {Mader}}{{Kochanek} et~al.}{2003}]{Koc03}
{Kochanek} C.~S.,  {White} M.,  {Huchra} J.,  {Macri} L.,  {Jarrett} T.~H.,
  {Schneider} S.~E.,   {Mader} J.,  2003, \mn@doi [\apj] {10.1086/345896},
  \href {http://adsabs.harvard.edu/abs/2003ApJ...585..161K} {585, 161}

\bibitem[\protect\citeauthoryear{{Lavaux} \& {Wandelt}}{{Lavaux} \&
  {Wandelt}}{2012}]{LavWan12}
{Lavaux} G.,  {Wandelt} B.~D.,  2012, \mn@doi [\apj]
  {10.1088/0004-637X/754/2/109}, \href
  {http://adsabs.harvard.edu/abs/2012ApJ...754..109L} {754, 109}

\bibitem[\protect\citeauthoryear{{Lee} \& {Park}}{{Lee} \&
  {Park}}{2009}]{LeePar09}
{Lee} J.,  {Park} D.,  2009, \mn@doi [\apjl] {10.1088/0004-637X/696/1/L10},
  \href {http://adsabs.harvard.edu/abs/2009ApJ...696L..10L} {696, L10}

\bibitem[\protect\citeauthoryear{{Liu}, {Zhang}  \& {Parsons}}{{Liu}
  et~al.}{2016}]{Liu16}
{Liu} A.,  {Zhang} Y.,   {Parsons} A.~R.,  2016, \mn@doi [\apj]
  {10.3847/1538-4357/833/2/242}, \href
  {http://adsabs.harvard.edu/abs/2016ApJ...833..242L} {833, 242}

\bibitem[\protect\citeauthoryear{{Mar{\'{\i}}n}, {Gnedin}, {Seo}  \&
  {Vallinotto}}{{Mar{\'{\i}}n} et~al.}{2010}]{Mar10}
{Mar{\'{\i}}n} F.~A.,  {Gnedin} N.~Y.,  {Seo} H.-J.,   {Vallinotto} A.,  2010,
  \mn@doi [\apj] {10.1088/0004-637X/718/2/972}, \href
  {http://adsabs.harvard.edu/abs/2010ApJ...718..972M} {718, 972}

\bibitem[\protect\citeauthoryear{{McQuinn}, {Zahn}, {Zaldarriaga}, {Hernquist}
  \& {Furlanetto}}{{McQuinn} et~al.}{2006}]{McQ06}
{McQuinn} M.,  {Zahn} O.,  {Zaldarriaga} M.,  {Hernquist} L.,   {Furlanetto}
  S.~R.,  2006, \mn@doi [\apj] {10.1086/505167}, \href
  {http://adsabs.harvard.edu/abs/2006ApJ...653..815M} {653, 815}

\bibitem[\protect\citeauthoryear{{Melin}, {Bartlett}  \&
  {Delabrouille}}{{Melin} et~al.}{2006}]{MelBarDel06}
{Melin} J.-B.,  {Bartlett} J.~G.,   {Delabrouille} J.,  2006, \mn@doi [\aap]
  {10.1051/0004-6361:20065034}, \href
  {http://adsabs.harvard.edu/abs/2006A%26A...459..341M} {459, 341}

\bibitem[\protect\citeauthoryear{{Newburgh} et~al.,}{{Newburgh}
  et~al.}{2016}]{HIRAX}
{Newburgh} L.~B.,  et~al., 2016, in Society of Photo-Optical Instrumentation
  Engineers (SPIE) Conference Series. p. 99065X (\mn@eprint {arXiv}
  {1607.02059}), \mn@doi{10.1117/12.2234286}

\bibitem[\protect\citeauthoryear{{Overzier}}{{Overzier}}{2016}]{Ove16}
{Overzier} R.~A.,  2016, \mn@doi [\aapr] {10.1007/s00159-016-0100-3}, \href
  {http://adsabs.harvard.edu/abs/2016A%26ARv..24...14O} {24, 14}

\bibitem[\protect\citeauthoryear{{Padmanabhan} \& {Refregier}}{{Padmanabhan} \&
  {Refregier}}{2017}]{PadRef17}
{Padmanabhan} H.,  {Refregier} A.,  2017, \mn@doi [\mnras]
  {10.1093/mnras/stw2706}, \href
  {http://adsabs.harvard.edu/abs/2017MNRAS.464.4008P} {464, 4008}

\bibitem[\protect\citeauthoryear{{Padmanabhan}, {Choudhury}  \&
  {Refregier}}{{Padmanabhan} et~al.}{2015}]{Pad15}
{Padmanabhan} H.,  {Choudhury} T.~R.,   {Refregier} A.,  2015, \mn@doi [\mnras]
  {10.1093/mnras/stu2702}, \href
  {http://adsabs.harvard.edu/abs/2015MNRAS.447.3745P} {447, 3745}

\bibitem[\protect\citeauthoryear{{Pober}}{{Pober}}{2015}]{Pob15}
{Pober} J.~C.,  2015, \mn@doi [\mnras] {10.1093/mnras/stu2575}, \href
  {http://adsabs.harvard.edu/abs/2015MNRAS.447.1705P} {447, 1705}

\bibitem[\protect\citeauthoryear{{Pober} et~al.,}{{Pober}
  et~al.}{2013}]{BAObab13}
{Pober} J.~C.,  et~al., 2013, \mn@doi [\aj] {10.1088/0004-6256/145/3/65}, \href
  {http://adsabs.harvard.edu/abs/2013AJ....145...65P} {145, 65}

\bibitem[\protect\citeauthoryear{{Rao}, {Turnshek}  \& {Nestor}}{{Rao}
  et~al.}{2006}]{Rao06}
{Rao} S.~M.,  {Turnshek} D.~A.,   {Nestor} D.~B.,  2006, \mn@doi [\apj]
  {10.1086/498132}, \href {http://adsabs.harvard.edu/abs/2006ApJ...636..610R}
  {636, 610}

\bibitem[\protect\citeauthoryear{{Reid} \& {White}}{{Reid} \&
  {White}}{2011}]{ReiWhi11}
{Reid} B.~A.,  {White} M.,  2011, \mn@doi [\mnras]
  {10.1111/j.1365-2966.2011.19379.x}, \href
  {http://adsabs.harvard.edu/abs/2011MNRAS.417.1913R} {417, 1913}

\bibitem[\protect\citeauthoryear{{Reid}, {Seo}, {Leauthaud}, {Tinker}  \&
  {White}}{{Reid} et~al.}{2014}]{Rei14}
{Reid} B.~A.,  {Seo} H.-J.,  {Leauthaud} A.,  {Tinker} J.~L.,   {White} M.,
  2014, \mn@doi [\mnras] {10.1093/mnras/stu1391}, \href
  {http://adsabs.harvard.edu/abs/2014MNRAS.444..476R} {444, 476}

\bibitem[\protect\citeauthoryear{{Rood}}{{Rood}}{1988}]{Roo88}
{Rood} H.~J.,  1988, \mn@doi [\araa] {10.1146/annurev.aa.26.090188.001333},
  \href {http://adsabs.harvard.edu/abs/1988ARA%26A..26..245R} {26, 245}

\bibitem[\protect\citeauthoryear{{Rykoff} et~al.,}{{Rykoff}
  et~al.}{2014}]{Ryk14}
{Rykoff} E.~S.,  et~al., 2014, \mn@doi [\apj] {10.1088/0004-637X/785/2/104},
  \href {http://adsabs.harvard.edu/abs/2014ApJ...785..104R} {785, 104}

\bibitem[\protect\citeauthoryear{{S{\'a}nchez} et~al.,}{{S{\'a}nchez}
  et~al.}{2017}]{San17}
{S{\'a}nchez} C.,  et~al., 2017, \mn@doi [\mnras] {10.1093/mnras/stw2745},
  \href {http://adsabs.harvard.edu/abs/2017MNRAS.465..746S} {465, 746}

\bibitem[\protect\citeauthoryear{{Seehars}, {Paranjape}, {Witzemann},
  {Refregier}, {Amara}  \& {Akeret}}{{Seehars} et~al.}{2016}]{See16}
{Seehars} S.,  {Paranjape} A.,  {Witzemann} A.,  {Refregier} A.,  {Amara} A.,
  {Akeret} J.,  2016, \mn@doi [\jcap] {10.1088/1475-7516/2016/03/001}, \href
  {http://adsabs.harvard.edu/abs/2016JCAP...03..001S} {3, 001}

\bibitem[\protect\citeauthoryear{{Seo} \& {Hirata}}{{Seo} \&
  {Hirata}}{2016}]{SeoHir16}
{Seo} H.-J.,  {Hirata} C.~M.,  2016, \mn@doi [\mnras] {10.1093/mnras/stv2806},
  \href {http://adsabs.harvard.edu/abs/2016MNRAS.456.3142S} {456, 3142}

\bibitem[\protect\citeauthoryear{{Seo}, {Dodelson}, {Marriner}, {Mcginnis},
  {Stebbins}, {Stoughton}  \& {Vallinotto}}{{Seo} et~al.}{2010}]{Seo2010}
{Seo} H.-J.,  {Dodelson} S.,  {Marriner} J.,  {Mcginnis} D.,  {Stebbins} A.,
  {Stoughton} C.,   {Vallinotto} A.,  2010, \mn@doi [\apj]
  {10.1088/0004-637X/721/1/164}, \href
  {http://adsabs.harvard.edu/abs/2010ApJ...721..164S} {721, 164}

\bibitem[\protect\citeauthoryear{{Shaw}, {Sigurdson}, {Pen}, {Stebbins}  \&
  {Sitwell}}{{Shaw} et~al.}{2014}]{Sha14}
{Shaw} J.~R.,  {Sigurdson} K.,  {Pen} U.-L.,  {Stebbins} A.,   {Sitwell} M.,
  2014, \mn@doi [\apj] {10.1088/0004-637X/781/2/57}, \href
  {http://adsabs.harvard.edu/abs/2014ApJ...781...57S} {781, 57}

\bibitem[\protect\citeauthoryear{{Shaw}, {Sigurdson}, {Sitwell}, {Stebbins}  \&
  {Pen}}{{Shaw} et~al.}{2015}]{Sha15}
{Shaw} J.~R.,  {Sigurdson} K.,  {Sitwell} M.,  {Stebbins} A.,   {Pen} U.-L.,
  2015, \mn@doi [\prd] {10.1103/PhysRevD.91.083514}, \href
  {http://adsabs.harvard.edu/abs/2015PhRvD..91h3514S} {91, 083514}

\bibitem[\protect\citeauthoryear{{Solanes}, {Manrique},
  {Garc{\'{\i}}a-G{\'o}mez}, {Gonz{\'a}lez-Casado}, {Giovanelli}  \&
  {Haynes}}{{Solanes} et~al.}{2001}]{Sol01}
{Solanes} J.~M.,  {Manrique} A.,  {Garc{\'{\i}}a-G{\'o}mez} C.,
  {Gonz{\'a}lez-Casado} G.,  {Giovanelli} R.,   {Haynes} M.~P.,  2001, \mn@doi
  [\apj] {10.1086/318672}, \href
  {http://adsabs.harvard.edu/abs/2001ApJ...548...97S} {548, 97}

\bibitem[\protect\citeauthoryear{{Stark}, {White}, {Lee}  \& {Hennawi}}{{Stark}
  et~al.}{2015a}]{Sta15a}
{Stark} C.~W.,  {White} M.,  {Lee} K.-G.,   {Hennawi} J.~F.,  2015a, \mn@doi
  [\mnras] {10.1093/mnras/stv1620}, \href
  {http://adsabs.harvard.edu/abs/2015MNRAS.453..311S} {453, 311}

\bibitem[\protect\citeauthoryear{{Stark}, {Font-Ribera}, {White}  \&
  {Lee}}{{Stark} et~al.}{2015b}]{Sta15b}
{Stark} C.~W.,  {Font-Ribera} A.,  {White} M.,   {Lee} K.-G.,  2015b, \mn@doi
  [\mnras] {10.1093/mnras/stv1868}, \href
  {http://adsabs.harvard.edu/abs/2015MNRAS.453.4311S} {453, 4311}

\bibitem[\protect\citeauthoryear{{Switzer} et~al.,}{{Switzer}
  et~al.}{2013}]{Swi13}
{Switzer} E.~R.,  et~al., 2013, \mn@doi [\mnras] {10.1093/mnrasl/slt074}, \href
  {http://adsabs.harvard.edu/abs/2013MNRAS.434L..46S} {434, L46}

\bibitem[\protect\citeauthoryear{{Thompson}, {Moran}  \& {Swenson}}{{Thompson}
  et~al.}{2017}]{TMS}
{Thompson} A.~R.,  {Moran} J.~M.,   {Swenson} Jr. G.~W.,  2017, {Interferometry
  and Synthesis in Radio Astronomy, 3rd Edition}

\bibitem[\protect\citeauthoryear{{Tinker} \& {Conroy}}{{Tinker} \&
  {Conroy}}{2009}]{TinCon09}
{Tinker} J.~L.,  {Conroy} C.,  2009, \mn@doi [\apj]
  {10.1088/0004-637X/691/1/633}, \href
  {http://adsabs.harvard.edu/abs/2009ApJ...691..633T} {691, 633}

\bibitem[\protect\citeauthoryear{{Vanderlinde} \& {Chime
  Collaboration}}{{Vanderlinde} \& {Chime Collaboration}}{2014}]{CHIME14}
{Vanderlinde} K.,  {Chime Collaboration} 2014, in Exascale Radio Astronomy.

\bibitem[\protect\citeauthoryear{{Villaescusa-Navarro}
  et~al.,}{{Villaescusa-Navarro} et~al.}{2016}]{Vil16}
{Villaescusa-Navarro} F.,  et~al., 2016, \mn@doi [\mnras]
  {10.1093/mnras/stv2904}, \href
  {http://adsabs.harvard.edu/abs/2016MNRAS.456.3553V} {456, 3553}

\bibitem[\protect\citeauthoryear{{White}}{{White}}{2002}]{TreePM}
{White} M.,  2002, \mn@doi [\apjs] {10.1086/342752}, \href
  {http://adsabs.harvard.edu/abs/2002ApJS..143..241W} {143, 241}

\bibitem[\protect\citeauthoryear{{White}, {Carlstrom}, {Dragovan}  \&
  {Holzapfel}}{{White} et~al.}{1999}]{Whi99}
{White} M.,  {Carlstrom} J.~E.,  {Dragovan} M.,   {Holzapfel} W.~L.,  1999,
  \mn@doi [\apj] {10.1086/306911}, \href
  {http://adsabs.harvard.edu/abs/1999ApJ...514...12W} {514, 12}

\bibitem[\protect\citeauthoryear{{White}, {Reid}, {Chuang}, {Tinker},
  {McBride}, {Prada}  \& {Samushia}}{{White} et~al.}{2015}]{Whi15}
{White} M.,  {Reid} B.,  {Chuang} C.-H.,  {Tinker} J.~L.,  {McBride} C.~K.,
  {Prada} F.,   {Samushia} L.,  2015, \mn@doi [\mnras] {10.1093/mnras/stu2460},
  \href {http://adsabs.harvard.edu/abs/2015MNRAS.447..234W} {447, 234}

\bibitem[\protect\citeauthoryear{{Wolz}, {Blake}  \& {Wyithe}}{{Wolz}
  et~al.}{2017}]{Wol17}
{Wolz} L.,  {Blake} C.,   {Wyithe} J.~S.~B.,  2017, preprint, \href
  {http://adsabs.harvard.edu/abs/2017arXiv170308268W} {} (\mn@eprint {arXiv}
  {1703.08268})

\bibitem[\protect\citeauthoryear{{Zaldarriaga}, {Furlanetto}  \&
  {Hernquist}}{{Zaldarriaga} et~al.}{2004}]{ZalFurHer04}
{Zaldarriaga} M.,  {Furlanetto} S.~R.,   {Hernquist} L.,  2004, \mn@doi [\apj]
  {10.1086/386327}, \href {http://adsabs.harvard.edu/abs/2004ApJ...608..622Z}
  {608, 622}

\bibitem[\protect\citeauthoryear{{Zhu}, {Pen}, {Yu}  \& {Chen}}{{Zhu}
  et~al.}{2016}]{ZhuTidal16}
{Zhu} H.-M.,  {Pen} U.-L.,  {Yu} Y.,   {Chen} X.,  2016, preprint, \href
  {http://adsabs.harvard.edu/abs/2016arXiv161007062Z} {} (\mn@eprint {arXiv}
  {1610.07062})

\bibitem[\protect\citeauthoryear{{van de Weygaert} \& {Platen}}{{van de
  Weygaert} \& {Platen}}{2011}]{Wey11}
{van de Weygaert} R.,  {Platen} E.,  2011, \mn@doi [International Journal of
  Modern Physics Conference Series] {10.1142/S2010194511000092}, \href
  {http://adsabs.harvard.edu/abs/2011IJMPS...1...41V} {1, 41}

\makeatother
\end{thebibliography}

\label{lastpage}
\end{document}